\documentclass[12pt]{article}%
\usepackage{comment}
\usepackage[nosort]{cite}
\usepackage{graphicx}
\usepackage{multicol}
\usepackage{amsfonts}
\usepackage{amssymb}
\usepackage{amsmath}
\usepackage{heck}
\usepackage[margin=1in]{geometry}
\usepackage{todonotes}
\usepackage{afterpage}
\usepackage{setspace}
\usepackage{verbatim}
\usepackage{color}
\usepackage{longtable}
\usepackage{float}
\usepackage{subcaption}
\usepackage{epsfig}
\usepackage{epstopdf}
\usepackage{adjustbox}
\usepackage{tikz}
\usepackage{titletoc}
\usepackage{hyperref}%
\usepackage{mathrsfs}
\usepackage{dsfont}
\usepackage{algorithm}
\usepackage{algpseudocode}
\usepackage{tikz}
\usepackage{pgfplots}
\usepackage{algorithmicx, algpseudocode,algorithm}
\usepackage{caption}
\usepackage{float}

\numberwithin{equation}{section}
\newsavebox{\mysavebox}

\hypersetup{colorlinks,citecolor=black,filecolor=black,linkcolor=black,urlcolor=black}

\usetikzlibrary{decorations.markings}
\usetikzlibrary{decorations.markings}
\usetikzlibrary{hobby}
\usepackage{tikz}
\usetikzlibrary{arrows}
\usetikzlibrary{arrows.meta}
\usetikzlibrary{arrows,decorations.pathmorphing}
\usetikzlibrary{shapes.geometric,calc,arrows, positioning,shapes.misc,decorations.markings,decorations.pathmorphing,arrows.meta}
\tikzset{
  big arrow/.style={
    decoration={markings,mark=at position 1 with {\arrow[scale=2,#1]{>}}},
    postaction={decorate},
    shorten >=0.4pt},
  big arrow/.default=black}

\pgfdeclarelayer{edgelayer}
\pgfdeclarelayer{nodelayer}
\pgfsetlayers{edgelayer,nodelayer,main}
\pgfplotsset{compat=1.16}
\tikzstyle{none}=[inner sep=0pt]

\tikzstyle{NodeCross}=[draw, shape=circle, cross out, inner sep=0pt, minimum size=6pt,line width=0.25mm]
\tikzstyle{Circle}=[draw, shape=circle, black, fill=black, inner sep=0pt, minimum size=6pt]
\tikzstyle{circle}=[draw, shape=circle, black, fill=black, inner sep=0pt, minimum size=16pt]
\tikzstyle{Star}=[draw, shape=star, fill=black, star points=8, inner sep=0pt, minimum size=8pt]
\tikzstyle{CircleRed}=[draw, shape=circle, black, fill=red, inner sep=0pt, minimum size=6pt]
\tikzstyle{StarP}=[draw={rgb,255: red,128; green,0; blue,128}, shape=star, fill={rgb,256: red,128; green,0; blue,128}, star points=8, inner sep=0pt, minimum size=12pt]
\tikzstyle{ShadedCircRed}=[draw=red, shape=circle, fill={rgb, 255: red,255; green,114; blue, 118}, inner sep=0pt, minimum size=80pt, line width=0.5mm, fill opacity=0.2]
\tikzstyle{ShadedCircRed2}=[draw=red, shape=circle, fill={rgb, 255: red,255; green,114; blue, 118}, inner sep=0pt, minimum size=10pt]
\tikzstyle{ShadedCircRed3}=[draw=black, shape=rectangle, fill={rgb, 255: red,255; green,114; blue, 118}, inner sep=0pt, minimum size=113pt, line width=0.25mm]
\tikzstyle{ShadedCirc}=[draw=red, shape=circle, fill=white, inner sep=0pt, minimum size=45pt,  fill opacity=1.0,  line width=0.5mm]
\tikzstyle{CircleBlue}=[draw, shape=circle, fill=blue, inner sep=0pt, minimum size=6pt]
\tikzstyle{BigCirclePurple}=[draw, shape=circle, fill={rgb,255: red,191; green,0; blue,191}, inner sep=0pt, minimum size=12pt]
\tikzstyle{CirclePurple}=[draw, shape=circle, fill={rgb,255: red,191; green,0; blue,191}, inner sep=0pt, minimum size=5pt]
\tikzstyle{EmptyCircle}=[draw, shape=circle, inner sep=0pt, minimum size=4pt]
\tikzstyle{GreenCircle}=[draw, shape=circle,  fill={rgb,255: red,80; green,200; blue,120}, inner sep=0pt, minimum size=8pt]
\tikzstyle{BrownCircle}=[draw, shape=circle,  fill={rgb,255: red,210; green,105; blue,30}, inner sep=0pt, minimum size=8pt]
\tikzstyle{CirclePurpleSmall}=[draw, shape=circle, fill={rgb,255: red,191; green,0; blue,191}, inner sep=0pt, minimum size=4pt]
\tikzstyle{BigCircleGreen}=[draw, shape=circle, fill={rgb,255: red,0; green,191; blue,0}, inner sep=0pt, minimum size=12pt]
\tikzstyle{BigCircleBlue}=[draw, shape=circle, fill={rgb,255: red,0; green,0; blue,191}, inner sep=0pt, minimum size=12pt]
\tikzstyle{BigCircleRed}=[draw, shape=circle, fill={rgb,255: red,191; green,0; blue,0}, inner sep=0pt, minimum size=12pt]
\tikzstyle{BrownCircleSmall}=[draw, shape=circle,  fill={rgb,255: red,210; green,105; blue,30}, inner sep=0pt, minimum size=6pt]
\tikzstyle{SmallCircleBrown}=[draw, shape=circle,  fill={rgb,255: red,210; green,105; blue,30}, inner sep=0pt, minimum size=5pt]
\tikzstyle{SmallCircleRed}=[draw, shape=circle, fill={rgb,255: red,191; green,0; blue,0}, inner sep=0pt, minimum size=6pt]
\tikzstyle{DashedLine}=[-, densely dashed, line width=0.25mm]
\tikzstyle{DottedLine}=[-, dotted, line width=0.25mm]
\tikzstyle{ThickLine}=[-, line width=0.25mm]
\tikzstyle{ArrowLineRight}=[-, -{Stealth[scale=1.25]}, line width=0.25mm, scale=5]
\tikzstyle{ArrowLineRed}=[-, draw={rgb,255: red,191; green,0; blue,0}, -{Stealth[scale=1.75]}, line width=0.1mm, scale=5]
\tikzstyle{RedLine}=[-, draw={rgb,255: red,191; green,0; blue,0}, fill=none, line width=0.5mm]
\tikzstyle{DashedLineThin}=[-, densely dashed, line width=0.125mm, fill=none, draw=black]
\tikzstyle{DottedRed}=[-, dotted, draw={rgb,255: red,191; green,0; blue,0}, fill=none, line width=0.25mm]
\tikzstyle{DashedRed}=[-, densely dashed, draw={rgb,255: red,191; green,0; blue,0}, fill=none, line width=0.25mm]
\tikzstyle{BlueLine}=[-, draw={rgb,255: red,0; green,0; blue,191}, fill=none, line width=0.5mm]
\tikzstyle{ArrowLineBlue}=[-, draw={rgb,255: red,0; green,0; blue,191}, -{Stealth[scale=1.75]}, line width=0.1mm, scale=5]
\tikzstyle{GreenDoubleArrow}=[<->, draw={rgb,155: red,0; green,255; blue,0},  line width= 0.5mm, scale=5]
\tikzstyle{RedDoubleArrow}=[<->, draw={rgb,255: red,255; green,0; blue,0},  line width= 0.5mm, scale=5]
\tikzstyle{BlueDottedLight}=[-, dotted, draw={rgb,255: red,0; green,0; blue,191}, fill=none, line width=0.3mm]
\tikzstyle{BrownLine}=[-, draw={rgb,255: red,210; green,105; blue,30}, fill=none, line width=0.5mm]
\tikzstyle{DottedRed}=[-, dotted, draw={rgb,255: red,191; green,0; blue,0}, fill=none, dotted, line width=0.5mm]
\tikzstyle{DottedPurple}=[-, dotted, draw={rgb,255: red,191; green,0; blue,191}, fill=none, dotted, line width=0.5mm]
\tikzstyle{BlueDottedLight}=[-, dotted, draw={rgb,255: red,0; green,0; blue,191}, fill=none, line width=0.5mm]
\tikzstyle{ArrowLinePurple}=[-, draw={rgb,255: red,191; green,0; blue,191}, -{Stealth[scale=1.75]}, line width=0.5mm, scale=5]
\tikzstyle{DashedLineGreen}=[-, densely dashed, draw={rgb,255: red,74; green,103; blue,65}, line width=0.25mm]
\tikzstyle{LineGreen}=[-, draw={rgb,255: red, 74; green,200; blue,65}, line width=0.5mm]
\tikzstyle{ArrowLineGreen}=[-, draw={rgb,255: red,0; green,191; blue,0}, -{Stealth[scale=1.75]}, line width=0.5mm, scale=5]
\tikzstyle{GreenLine}=[-, draw={rgb,255: red,0; green,191; blue,0}, fill=none, line width=0.5mm]
\tikzstyle{PurpleLine}=[-, draw={rgb,255: red,191; green,0; blue,191}, fill=none, line width=0.5mm]
\tikzstyle{PPurpleLine}=[-, draw={rgb,255: red,191; green,0; blue,191}, fill=none, line width=2.5mm]
\tikzstyle{DPurpleLine}=[-, dotted, draw={rgb,255: red,191; green,0; blue,191}, fill=none, line width=0.5mm]
\tikzstyle{SBrownLine}=[-, draw={rgb,255: red,191; green,0; blue,191}, fill=none, opacity=0.35, line width=2.5mm]
\tikzstyle{DottedBlue}=[-, dotted, draw=blue, fill=none, dotted, line width=0.5mm]
\tikzstyle{DashedPurpleLine}=[-, densely dashed, draw={rgb,255: red,191; green,0; blue,191}, fill=none, line width=0.5mm]
\tikzstyle{SmallCircleBlue}=[draw, shape=circle, fill=blue, inner sep=0pt, minimum size=5pt]
\tikzstyle{SmallCirclePurple}=[draw, shape=circle, fill={rgb,255: red,191; green,0; blue,191}, inner sep=0pt, minimum size=5pt]
\tikzset{snake it/.style={decorate, decoration=snake}}

\tikzset{
dashstar/.style={
 dash pattern=on 5pt off 5pt,
 postaction={
  decorate,
  decoration={
   markings,
   mark=between positions 9pt and 1 step 10pt with {
     \node[color=red] {*};
   }
  }
 }
},
dashstarstar/.style={ 
 dash pattern=on 5pt off 10pt,
 postaction={
   decorate,
   decoration={
     markings,
     mark=between positions 10pt and 1
          step 15pt
           with {
            \node at (-2pt,0pt) {\pgfuseplotmark{star}};
            \node at (2pt,0pt) {\pgfuseplotmark{star}};
           }
   }
 }
}
}

\begin{document}

\date{November 2026}

\title{On Boundary State Screening via Solitons}

\institution{PENN}{\centerline{$^{1}$Department of Physics and Astronomy, University of Pennsylvania, Philadelphia, PA 19104, USA}}
\institution{YAUmath}{\centerline{$^{2}$Yau Mathematical Sciences Center, Tsinghua University, Beijing, 100084, China}}
\institution{BEIJINGmath}{\centerline{$^{3}$Beijing Institute of Mathematical Sciences and Applications, Huairou District, Beijing, 101408, China}}
\institution{PENNmath}{\centerline{$^{4}$Department of Mathematics, University of Pennsylvania, Philadelphia, PA 19104, USA}}

\authors{
Vivek Chakrabhavi\worksat{\PENN}\footnote{e-mail: \texttt{vivekcm@sas.upenn.edu}},
Babak Haghighat\worksat{\YAUmath,\BEIJINGmath}\footnote{e-mail: \texttt{babakhaghighat@tsinghua.edu.cn}},
and Jonathan J. Heckman\worksat{\PENN,\PENNmath}\footnote{e-mail: \texttt{jheckman@sas.upenn.edu}}
}

\abstract{
We present a coarse classification scheme for the boundary data of massive QFTs based on transitions of boundary states via soliton / kink profiles. Starting from the lattice of boundary states, we introduce the boundary screening group $\mathbb{B}$: the equivalence classes of boundary states which cannot be connected by such transitions. In string theory backgrounds which geometrically engineer an interacting QFT in the target space, $\mathbb{B}$ computes the worldsheet analog of the target space defect group $\mathbb{D}$. When $\mathbb{B}$ is non-zero, it provides a diagnostic for having a target space with non-normalizable modes. For interfaces between QFTs, similar considerations imply the existence of a non-trivial conserved charge associated with soliton absorption / transmission, which means that the interface cannot fully dissolve / disappear. Similar considerations hold for junctions of QFTs, which in the special case of string worldsheet CFTs translate to junctions of string theories. We illustrate these general considerations with a number of non-geometric 2D $\mathcal{N} = (2,2)$ Landau-Ginzburg models, as well as their natural lifts to 4D $\mathcal{N} = 1$ Wess-Zumino models.
}

\maketitle

\enlargethispage{\baselineskip}

\setcounter{tocdepth}{3}

\tableofcontents

\newpage

\section{Introduction} \label{sec:INTRO}

While it is sometimes stated that science is a differential equation and religion is a boundary condition,\footnote{Attributed, as far as we are aware, to Alan Turing.} boundary conditions play an important role in the study of many quantum field theories. This includes properly defining scattering amplitudes, time evolution of states, as well as many properties of bulk / boundary systems. Giving a complete characterization of such boundary data is an important but challenging task.

Given a $D$-dimensional QFT with one direction on an interval, fully specifying the theory requires a choice of boundary conditions $B_\alpha$, $B_\beta$ at the two ends. In the case of a QFT with massive vacua, soliton profiles specify interpolating field profiles connecting these boundary conditions. Letting $\Lambda$ denote the lattice spanned by the corresponding
boundary states $\vert B_\alpha \rangle$, these solitons determine a canonical intersection 
pairing $\Lambda \times \Lambda \rightarrow \mathbb{Z}$,
as well as a linear map:
\begin{equation}
\mu: \Lambda \rightarrow \Lambda.
\end{equation}
This defines a collection of transitions between boundary states via (signed) kink profiles. 
In this paper we ask which states cannot be connected by such transitions.
This is captured by the quotient:
\begin{equation}
\mathbb{B} = \Lambda / \mathrm{im} (\mu),
\end{equation}
which we refer to as the boundary state screening group (BSSG).

As defined, this quantity makes sense for any massive QFT in which the boundary states fill out a $\mathbb{Z}$-module with a canonical pairing (as dictated by a signed partition function) i.e., it applies to massive theories and their CFT counterparts. It also naturally arises in a number of string-based constructions of geometrically engineered QFTs.
From the target space viewpoint, the boundary states of interest are associated with D-branes wrapped on non-compact cycles (i.e., non-normalizable states) as well as compact cycles (i.e., normalizable states). There is a natural grading for the D-branes specified by their RR charge \cite{Polchinski:1995mt}. From the worldsheet perspective, each D-brane corresponds to a boundary state for an open string, and the quotient $\mathbb{B} = \Lambda / \mathrm{im}(\mu)$ measures the same information computed by the target space defect group \cite{DelZotto:2015isa}. In the context of string theory one can also contemplate non-geometric target spaces, i.e., the worldsheet theory is specified by a more general CFT. Canonical examples include Landau-Ginzburg models, as well as asymmetric orbifolds.\footnote{There is a large body of work on the spectrum of boundary states / D-branes in LG models, though as far as we are aware, a similar systematic study does not exist for asymmetric orbifolds.} In these cases, $\mathbb{B}$ captures the same structure of screening of states in the target space.

The boundary state screening group $\mathbb{B}$ provides access to the structure of the QFT itself. For one, if it is non-zero, this serves as a coarse diagnostic that the QFT has a non-compact target space, i.e., it contains non-normalizable states. Additionally, there can be interfaces between distinct QFTs of the form:
\begin{equation}
\mathcal{T}_1 \vert {\mathbf{I}} \vert \mathcal{T}_2.
\end{equation}
These can be viewed as boundary
conditions of a larger theory, i.e., of the folded theory $\mathcal{T}_1 \otimes \overline{\mathcal{T}}_2$ with $\overline{\mathcal{T}}_2$ the
CRT conjugate. In this enlarged theory, we can also compute a boundary state screening group, and the interface can be assigned a non-trivial conserved charge. The presence of this charge indicates that some solitons cannot be transmitted across, and that the interface itself cannot completely dissolve / disappear in the target space. Said differently, this provides a worldsheet diagnostic for the stability of domain walls in the target space. The existence of such domain walls is expected from the Swampland cobordism conjecture \cite{McNamara:2019rup}, and their stability hinges on whether the accompanying bordism group is non-trivial, i.e., it cannot be smoothed out by any field configuration in the low energy effective field theory \cite{Kaidi:2024cbx, Heckman:2025wqd}.

Similar considerations hold for junctions of multiple QFTs all joined at a common interface. In the case of string worldsheet theories, this
corresponds to a characterization of junctions of distinct perturbative string theories, as in
\cite{Altavista:2026edv, Altavista:2026muc, Tachikawa:2026jaj}. The presence
of a conserved boundary state screening group charge means that the junction in the target space cannot dissolve / disappear.

We illustrate these general points with some well-known 2D examples where the target space does not have an immediate ``geometric interpretation.'' The main class of examples we consider consists of Landau-Ginzburg models. These are massive QFTs but provide a
characterization of ``non-geometric'' phases of a string theory background. In this case, the relevant spectrum of BPS D-brane states is encoded in a matrix of solitons between massive vacua \cite{Hori:2000ck} (see also \cite{Gaiotto:2015aoa, Gaiotto:2015zna}). The construction naturally generalizes to higher-dimensional QFTs, and we illustrate these points for 4D $\mathcal{N} = 1$ Wess-Zumino models, i.e., the parent theories for 2D $\mathcal{N} = (2,2)$ Landau-Ginzburg models. We anticipate that similar considerations hold for general QFTs.

The rest of this paper is organized as follows. In section \ref{sec:SCREEN} we introduce the boundary soliton screening group, and in
section \ref{sec:EXAMPLES} we present some illustrative examples involving massive LG models. In section \ref{sec:INTERFACE} we
use the boundary screening group to study properties of interfaces between QFTs. We briefly discuss the generalization to junctions of theories in section \ref{sec:JUNCTION}. We discuss the generalization to higher-dimensional massive QFTs in section \ref{sec:HIGHER}. We present our conclusions and potential future directions in section \ref{sec:CONC}. In Appendix \ref{app:LG} we review some general properties of D-branes in LG models. We review additional details on interfaces in Landau-Ginzburg theories in Appendix \ref{app:LGinterface}. We present a non-trivial compact example with trivial $\mathbb{B}$ in Appendix \ref{app:FURTHER}.

\section{Boundary State Screening} \label{sec:SCREEN}

In this section we introduce the boundary state screening group, focusing for ease of exposition on the case of 2D quantum field theories (QFTs).\footnote{We emphasize, however, that we expect most of the construction to carry over to higher-dimensional settings as well. See section \ref{sec:HIGHER} for some discussion along these lines.}
The main idea will be to consider a 2D QFT on the spacetime $\mathbb{R}_{\mathrm{time}} \times I^{\mathrm{space}}_{\alpha \beta}$, i.e., a QFT on the interval with boundary conditions $B_\alpha$ and $B_\beta$ on two sides of an interval. We shall also find it helpful to consider the related Euclidean setup on the cylinder $S^{1}_{\mathrm{time}} \times I^{\mathrm{space}}_{\alpha \beta}$ as well as the channel-dual description with $I^{\mathrm{time}}_{\alpha \beta} \times S^{1}_{\mathrm{space}}$ characterizing a transition amplitude. We shall be interested in situations where the boundary conditions of interest span a finite-dimensional subspace of the full Hilbert space of states. This is the situation in QFTs with massive vacua, as well as in
situations with a finite spectrum of BPS boundary conditions.

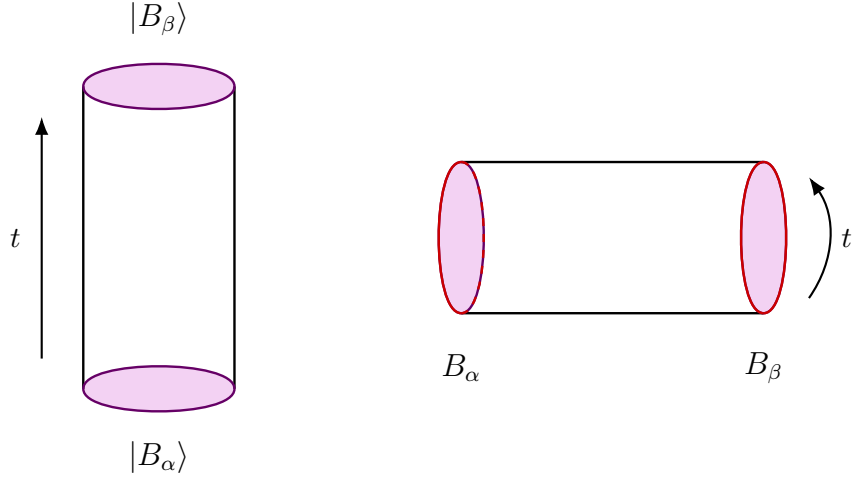
\begin{figure}[t!]
    \centering

\begin{tikzpicture}[x=1cm,y=1cm,
  none/.style={inner sep=0pt, outer sep=0pt},
  ThickLine/.style={draw=black, line width=0.9pt},
  PurpleLine/.style={draw=violet!80!black, line width=0.9pt},
  ArrowLineRight/.style={
    -{Latex[length=2.4mm,width=1.8mm]}, line width=0.8pt},
  RedLine/.style={draw=red!80!black, line width=0.9pt},
  DashedRed/.style={draw=red!80!black, line width=0.9pt, dashed}
]

  \begin{pgfonlayer}{nodelayer}
    \node [style=none] (aTL) at (1, 2) {};
    \node [style=none] (aTR) at (3, 2) {};
    \node [style=none] (aBL) at (1, -2) {};
    \node [style=none] (aBR) at (3, -2) {};
    \node [style=none] (aA0) at (0.45, -1.6) {};
    \node [style=none] (aA1) at (0.45, 1.6) {};

    \node [style=none] at (2, 2.90) {$\vert B_\beta \rangle$};
    \node [style=none] at (2, -2.90) {$|B_\alpha\rangle$};
    \node [style=none] at (0.1, 0) {\small $t$};

    \node [style=none] (oTL) at (6, 1) {};
    \node [style=none] (oBL) at (6, -1) {};
    \node [style=none] (oTR) at (10, 1) {};
    \node [style=none] (oBR) at (10, -1) {};
    \node [style=none] (oA0) at (10.6, -0.8) {};
    \node [style=none] (oA1) at (10.6, 0.8) {};
    \node [style=none] at (6, -1.72) {$B_\alpha$};
    \node [style=none] at (10, -1.72) {$B_\beta$};
    \node [style=none] at (11.1, 0) {\small $t$};
  \end{pgfonlayer}

  \begin{pgfonlayer}{edgelayer}
    \draw [style=ThickLine] (aTL.center) to (aBL.center);
    \draw [style=ThickLine] (aTR.center) to (aBR.center);
    \draw [style=PurpleLine,
      fill={rgb,255:red,191; green,0; blue,191}, fill opacity=0.18]
      (aTL.center) to [out=-90, in=-90, looseness=0.51] (aTR.center)
      to [out=90, in=90, looseness=0.51] (aTL.center);
    \draw [style=PurpleLine,
      fill={rgb,255:red,191; green,0; blue,191}, fill opacity=0.18]
      (aBL.center) to [out=-90, in=-90, looseness=0.51] (aBR.center)
      to [out=90, in=90, looseness=0.51] (aBL.center);
    \draw [style=ArrowLineRight] (aA0.center) to (aA1.center);

    \draw [style=ThickLine] (oTL.center) to (oTR.center);
    \draw [style=ThickLine] (oBL.center) to (oBR.center);

    \draw [style=PurpleLine,
      fill={rgb,255:red,191; green,0; blue,191}, fill opacity=0.18]
      (oTL.center) to [out=180, in=180, looseness=0.51] (oBL.center)
      to [out=0, in=0, looseness=0.51] (oTL.center);
    \draw [style=PurpleLine,
      fill={rgb,255:red,191; green,0; blue,191}, fill opacity=0.18]
      (oTR.center) to [out=180, in=180, looseness=0.51] (oBR.center)
      to [out=0, in=0, looseness=0.51] (oTR.center);

    \draw [style=RedLine]
      (oTL.center) to [out=180, in=180, looseness=0.51] (oBL.center);
    \draw [style=DashedRed]
      (oTL.center) to [out=0, in=0, looseness=0.51] (oBL.center);
    \draw [style=RedLine]
      (oTR.center) to [out=180, in=180, looseness=0.51] (oBR.center);
    \draw [style=RedLine]
      (oTR.center) to [out=0, in=0, looseness=0.51] (oBR.center);
    \draw [style=ArrowLineRight, bend right=35]
      (oA0.center) to (oA1.center);
  \end{pgfonlayer}
\end{tikzpicture}
    \caption{Depiction of the signed partition function with boundary states $\vert B_\alpha \rangle$ and $\vert B_{\beta} \rangle$ in the ``closed string'' channel, and its equivalent evaluation in terms of a signed partition function with boundary conditions $B_{\alpha}$ and $B_{\beta}$. Channel duality enforces the condition that the physically admissible boundary states form a $\mathbb{Z}$-module.}
    \label{fig:BOUNDARYSTATES}
\end{figure}

It is natural to consider the $\mathbb{Z}$-module of all possible boundary conditions, i.e., all possible integral linear combinations. The restriction to all possible integral linear combinations is a mild generalization of the Cardy consistency conditions for conformal boundary conditions in a 2D CFT (see \cite{Cardy:1989ir, Cardy:2004hm}). We can either evaluate the signed partition function i.e., allowing insertions by $(-1)^F$ and $(-1)^F F$ in the ``closed string channel'' or in the ``open string channel''. This imposes the restriction that we only deal with integral linear combinations of boundary states in the closed string channel and the corresponding $\mathbb{Z}$-module in the open string channel.\footnote{In a 2D CFT, we can start with conformal boundary conditions $B_{\alpha}$ and $B_{\beta}$ and evaluate a (possibly signed) partition function $\mathrm{Tr}_{\mathcal{H}_{\alpha \beta}} \varepsilon e^{-\beta H}$, where $\varepsilon $ can refer to either the identity, $(-1)^F$ or $(-1)^F F$. Going to the dual channel we can also evaluate $\langle B_{\alpha} \vert \varepsilon e^{-\beta H} \vert B_{\beta} \rangle$. Observe that in one channel we are simply counting signed states, i.e., possible integers. Equality of the two evaluations naturally means we restrict to integral linear combinations of boundary states. Our discussion here is a mild extension of the one presented in \cite{Collier:2021ngi} to include signs in related partition functions.} In what follows we shall abuse terminology and refer to $\vert B_{\alpha} \rangle$ as a boundary state, even when discussing the ``open string channel'' since it makes the structure of the $\mathbb{Z}$-module more manifest.

We denote by $\Lambda$ the integral lattice spanned by the states $\vert B_\alpha \rangle$.
Different boundary conditions are connected by kink / soliton profiles, which in the closed channel signal transitions:
\begin{equation}
\vert B_\alpha \rangle \rightarrow \vert B_\beta \rangle.
\end{equation}
This can be enlarged to formal transitions between integral linear combinations of these boundary states:
\begin{equation}\label{eq:BoundaryStateCombination}
\vert \chi \rangle = \underset{\alpha}{\sum} n^{(\chi)}_\alpha \vert B_\alpha \rangle.
\end{equation}

Equation~\eqref{eq:BoundaryStateCombination} describes a general, possibly non-elementary boundary condition in the additive framework used here. Its decomposition into elementary boundary states makes the corresponding transition amplitude a sum over elementary sectors, as discussed below. In a geometric string theory realization, these sectors correspond to D-brane configurations, and the total conserved charge is $q_\chi=\sum_\alpha n_\alpha^{(\chi)}q_\alpha$ in the appropriate D-brane K-theory group. Nonnegative coefficients describe brane multiplicities, while allowing signed coefficients extends this description to formal differences of charges. This additive charge description does not require the configuration to be an elementary or stable supersymmetric brane.

Indeed, we can always take the
path integral and break it up into a sum over transitions between vacua. 
Our interest will be in the spectrum of kink profiles / solitons connecting different boundary states. 
In particular, the solitons / kink profiles define a linear map:
\begin{equation}\label{eq:LINEAR}
\mu : \Lambda \rightarrow \Lambda.
\end{equation}
Given a boundary state $\vert \chi \rangle$, we ask whether there exists a corresponding boundary state $\vert \psi \rangle$ such that 
$\mu \vert \psi \rangle = \vert \chi \rangle$. The obstruction to the existence of such states defines a notion of partial 
screening for the space of boundary states.

Our proposal is motivated by the appearance of similar structures in the study of the generalized symmetries of QFTs engineered in string theory, i.e., the construction of a defect group \cite{Heckman:2015bfa} for the QFT.\footnote{We comment that the defect group is intrinsic data of the QFT and can be defined even when a stringy realization is not (yet) known.} The basic idea will be to start with a fixed choice $\vert \chi \rangle$ and construct all possible states in $\Lambda$
which can be realized via solitonic configurations. 
Formally speaking, we are interested in precisely those states which fail to be in the image of $\mu$, namely the cokernel:
\begin{equation}\label{eq:BSCREEN}
\mathbb{B} = \mathrm{coker}(\mu) = \Lambda / \mathrm{im}(\mu),
\end{equation}
where we refer to $\mathbb{B}$ as the ``boundary state screening group.''

In practice, $\mathbb{B}$ is a discrete group consisting of torsion and free factors:\footnote{We adhere to the conventions for non-compact models where the number of free factors is $f+1$, i.e., there is a continuous flavor symmetry of rank $f$ and a ``KK momentum'' degree of freedom which arises from compactification on a circle. In compact models one has $f + 1 = 0$.}
\begin{equation}
\mathbb{B} = \left(\underset{j}{\bigoplus} \mathbb{Z} / {p_j}^{e_j}\mathbb{Z} \right) \oplus \mathbb{Z}^{f+1}.
\end{equation}
When $\mathbb{B}$ is non-trivial, it tells us that there are boundary states which cannot be obtained as the endpoint of a transition. In terms of the corresponding target space, these can be viewed as defining an abstract global symmetry, though we emphasize that our discussion applies even when the target space of the QFT is completely non-geometric.

In string theory terms, having non-trivial $\mathbb{B}$ necessarily means that there are non-normalizable states in the target space.
From this perspective, $\mathbb{B}$ provides a simple diagnostic to check for non-compactness of the model.

The rest of this section is organized as follows. We begin by discussing a motivating example based on the defect group of
geometrically engineered QFTs in string theory. In this case, the relevant boundary states come from the spectrum of
BPS objects in string theory, and the boundary states in question are D-branes wrapped on cycles of a non-compact geometry.
This geometric setup serves as a starting point for our more abstract treatment to follow. With this in place, we next turn to the other extreme, as characterized by a 2D $\mathcal{N} = (2,2)$ Landau-Ginzburg theory with massive vacua.

\subsection{The Defect Group} \label{ssec:DEFECT}

Having given a general notion of the boundary state screening group, we now explain how it is related to the defect group of references \cite{DelZotto:2015isa, GarciaEtxebarria:2019caf, Albertini:2020mdx, Morrison:2020ool}. In this setting, one considers a QFT engineered via a string theory background in which the QFT degrees of freedom are localized at the tip of a topological cone $X$. Defects are realized via branes which stretch from the tip of the cone out to the boundary $\partial X$. These $p$-branes are charged under $(p+1)$-form gauge potentials which specify (upon compactification) candidate higher-form symmetries of the QFT which are partially screened by dynamical states obtained from branes wrapped on collapsing cycles near the tip of the singularity. Geometrically, this screening amounts to a defect in the QFT spacetime which can be terminated by a dynamical state. This partial screening involves a quotient of the form $H_{\ast}(X, \partial X) / H_{\ast}(X)$, i.e., a quotient of the relative homology group by compactly supported homology cycles. The overall global form of the QFT is specified by a choice of consistent boundary conditions, i.e., by specifying the spectrum of admissible branes in the first place.\footnote{There have been a number of refinements of this basic picture over the past few years, see e.g., \cite{DelZotto:2015isa, Albertini:2020mdx, Cvetic:2021maf, Antinucci:2022vyk, vanBeest:2022fss, Heckman:2022xgu, Bah:2023ymy, Apruzzi:2023uma, Cvetic:2023pgm, Baume:2023kkf, Yu:2023nyn, Heckman:2024oot, Bonetti:2024cjk, Braeger:2024jcj, Franco:2024mxa, Heckman:2024zdo, Cvetic:2024dzu, GarciaEtxebarria:2024jfv, Tian:2024dgl, Cvetic:2024mtt, Bonetti:2024etn, Heckman:2025lmw, Morrison:2020ool, Tian:2021cif, Braeger:2025rov, DelZotto:2022fnw, Heckman:2022muc, Bertolini:2025wyj, Cvetic:2022imb, Cvetic:2023plv, Acharya:2023bth, DelZotto:2024tae, DeMarco:2025pza, Dramburg:2025tlb, Chakrabhavi:2026iku, Harding:2026gzj, GarciaEtxebarria:2019caf, Apruzzi:2021nmk, Bhardwaj:2021mzl, Lawrie:2023tdz, Zhang:2024oas} for a partial list of references.} Schematically, the defect group takes the form:
\begin{equation}
\mathbb{D} = \underset{n}{\bigoplus} \mathbb{D}^{(n)},
\end{equation}
with:
\begin{equation}
\mathbb{D}^{(n)} = \underset{p-\mathrm{branes}}{\bigoplus}\,\, \underset{p + 1 = n + k}{\bigoplus} \frac{H_{k}(X, \partial X)}{H_{k}(X)}.
\end{equation}

For our purposes, the important point is that D-branes wrapped on cycles serve as boundary conditions for open strings, i.e., they specify boundary conditions $B_{\alpha}$ of the open string worldsheet CFT. Moreover, the ground states of open strings stretched between two such branes are, in a massive Landau-Ginzburg phase of the theory, simply solitons interpolating between different boundary conditions. From our brief review of the defect group, we see that branes can wrap non-compact cycles and thus define non-dynamical states in the QFT. Geometrically, these wrap cycles in the relative homology groups $H_{\ast}(X, \partial X)$, and these are partially screened by dynamical states obtained from the branes wrapped on the compact support homology groups $H_{\ast}(X)$.

In geometric settings, the basic data of the defect group can be read off from fairly minimal data of the background $X$. For example, when $X$ is a Calabi-Yau $n$-fold, we have a natural intersection pairing on middle-dimensional cycles which is anti-symmetric for $n$ odd and symmetric for $n$ even. This intersection form in turn specifies a Dirac pairing $\mu$ in the QFT spacetime \cite{DelZotto:2022fnw, DelZotto:2022ras, Braeger:2024jcj}. The corresponding defect group is then read off from the cokernel of $\mu$:
\begin{equation}\label{eq:DSCREEN}
\mathbb{D} = \mathrm{coker}( \mu ),
\end{equation}
which is simply the target space version of equation (\ref{eq:BSCREEN}). We comment that this same structure naturally appears in the BPS quiver for 4D KK theories obtained from the dimensional reduction of 5D SCFTs \cite{DelZotto:2022fnw, Tian:2021cif}, as well as 4D orbifold theories \cite{Braeger:2024jcj,Braeger:2025rov}. In that setting, $\mu$ is simply the anti-symmetrization of the adjacency matrix for the (fermionic) degrees of freedom of the accompanying quiver. This is in accord with the interpretation of the quiver as counting the ground states of strings stretched between branes. Strictly speaking, this correspondence with the Dirac pairing requires compactifications to a spacetime of dimension three or higher. In lower-dimensional systems, $\mu$ is the open string index, and the boundary state screening group $\mathbb{B}$ is related to the  K-theory of the target space geometry. This is closely related to, but a priori, distinct from the defect group constructed from relative (co)homology. Nevertheless, in many cases the corresponding (co)homological information can be recovered from the worldsheet index.

Summarizing, the boundary state screening group generalizes this basic formulation to more general QFTs which
include both non-geometric worldsheet theories, as well as more general 2D QFTs.

\subsection{Landau-Ginzburg Theories} \label{ssec:LG}

As already mentioned, the defect group is naturally defined in situations where we have a geometric target space. In many situations of interest, however, the 2D QFT may not admit a simple geometric characterization. We nevertheless expect the general considerations based on the boundary state screening group to still apply. As a particular example of this sort, we now discuss 2D $\mathcal{N} = (2,2)$ Landau-Ginzburg models. These are well-known to be connected to geometric sigma models (see e.g., \cite{Witten:1993yc}), and also often admit geometric mirror descriptions \cite{Hori:2000kt}. Our aim here will be to formulate the boundary screening group for this class of examples. Subsequent sections will present illustrative examples and applications.

Recall that a 2D Landau-Ginzburg (LG) model can be formulated in terms of 2D $\mathcal{N} = (2,2)$ chiral superfields $X_i$ (as obtained from dimensional reduction of 4D $\mathcal{N} = 1$ chiral superfields) and a corresponding superpotential $W(X_i)$. We shall be interested in LG models with a finite number of massive vacua, as specified by critical points of the superpotential, i.e., $\partial W|_{X = v} = 0$. These models have solitons connecting the different vacua, and the total number is calculated by a soliton matrix $N_{\alpha \beta}$ connecting the vacua $\alpha \rightarrow \beta$.

The count of solitons connecting vacua $\alpha$ and $\beta$ is controlled by the index \cite{Cecotti:1992rm}:
\begin{equation}
\vert \mu_{\alpha \beta} \vert = \left\vert \underset{\alpha \beta \, \mathrm{solitons}}{\sum} (-1)^F F \right\vert = \left\vert \underset{\alpha \beta \, \mathrm{bosonic \, solitons}}{\sum} (-1)^F \right\vert.
\end{equation}

Moreover, as explained in \cite{Cecotti:1992rm}, this implicitly defines an intersection pairing for the solitons; in a model with $n$ fields
the intersection pairing satisfies:
\begin{equation}\label{eq:SIGN}
\Delta_{\alpha} \circ \Delta_{\beta} = (-1)^{n-1} \Delta_{\beta} \circ \Delta_{\alpha}.
\end{equation}
Indeed, given $W_\alpha$ a critical point of $W$, observe the field profile of the soliton defines a topological $S^{n-1}$.\footnote{To see this, consider the 2D QFT on $\mathbb{R}_{\mathrm{time}} \times \mathbb{R}_{\sigma}$, with a vacuum configuration specified at $\sigma \rightarrow \pm \infty$. Observe that without loss of generality, we can approximate $W$ near the critical point as:
\begin{equation}
W  = \frac{1}{2} \underset{i}{\sum} X_i^2  .
\end{equation}
In a suitable normalization for the kinetic terms of the fields, the soliton equation (see Appendix \ref{app:LG} for a less schematic discussion) satisfies:
\begin{equation}
\frac{d X^{i}}{d \sigma} = \overline{X}^{i},
\end{equation}
with solution $X^{i} = \zeta^{i} e^{\sigma}$ as $\sigma \rightarrow - \infty$. The trajectory of the soliton specifies a straight line in the $W$-plane. Observe that for $r$ real, the profile $W = r^2$ specifies an $S^{n-1}$, namely a vanishing cycle. The sign of the intersection form appearing in equation (\ref{eq:SIGN}) then follows. Gauging with respect to a discrete symmetry changes the statistics of the solitons since they can now be dressed by a non-trivial Wilson line. See reference \cite{Cecotti:1992rm} for additional details on the statistics of such solitons. \label{foot:STAT}}
In addition to $\mu_{\alpha \beta}$, we shall also be interested in the directed
soliton matrix $N_{\alpha \beta}$ and the corresponding Stokes matrix $S = \mathbf{1}+N$ which connects two vacua. The relation between $\mu$ and $S$ is:
\begin{equation}\label{eq:muSYMM}
\mu = S + (-1)^{n-1} S^{T},
\end{equation}
where $S^{T}$ is the transpose of $S$. Equivalently, the intersection pairing $\mu_{\alpha\beta}$ is the open-string index between boundary conditions $B_{\alpha}$ and $B_{\beta}$,
\begin{equation}
I_{\alpha\beta} = \textrm{Tr}_{\mathcal{H}_{\alpha\beta}}(-1)^F F .
\end{equation}
In the Landau-Ginzburg description, this index computes the signed intersection number of the corresponding D-branes / vanishing cycles \cite{Hori:2000ck, Cecotti:1992rm, Feng:2002kk} (see also \cite{Gaiotto:2015aoa, Gaiotto:2015zna}).

\section{Illustrative Examples} \label{sec:EXAMPLES}

In this section we present some illustrative examples. We primarily focus on 2D $\mathcal{N} = (2,2)$ massive LG models.
We present both an intrinsic characterization of the accompanying boundary screening group, and when available compare it
to a dual geometric description.

We begin with $\mathcal{A}$- and $\mathcal{D}$-type LG minimal models.\footnote{There is an unfortunate clash of terminology in terms of the
minimal CFTs, the corresponding minimal LG models, and their relations. The context should make it clear which sort of ``minimal model'' we are referring to at a given time.} We then consider the LG models of relevance for the geometrically engineered $(A,D)$ Argyres-Douglas theories.
We also consider LG models for the duals of toric threefolds, where we have a geometric cross-check, and finally LG models for the duals of non-supersymmetric orbifolds.

\subsection{\texorpdfstring{$\mathcal{A}$}{A}-type Models} \label{app:ATypeMinimalModel}

In this section we consider the level-$k$ $\mathcal{A}$-type minimal model, namely the infrared fixed point of the Landau-Ginzburg theory
\begin{equation}
    W_0(X) = \frac{1}{k+2}X^{k+2}.
\end{equation}
The superpotential has a single degenerate critical point at $X=0$.  In order to resolve it into isolated massive vacua and their solitons, we introduce the relevant deformation
\begin{equation}
    W(X) = \frac{1}{k+2}X^{k+2}-X. \label{eq:DeformedAType}
\end{equation}

The critical-point equation is
\begin{equation}
    \frac{dW}{dX} = X^{k+1}-1 = 0.
\end{equation}
Consequently, the massive theory has $k+1$ isolated vacua,
\begin{equation}
    X_\alpha = \exp\left(\frac{2\pi i \alpha}{k+1}\right), \qquad \alpha=0,\ldots,k.
\end{equation}
At a critical point, $X_\alpha^{k+1}=1$, and hence the corresponding critical value is
\begin{equation}
    w_\alpha = W(X_\alpha) = -\frac{k+1}{k+2}X_\alpha.
\end{equation}
The critical values therefore lie at equally spaced points on a circle in the $W$-plane.

For the deformation \eqref{eq:DeformedAType}, there is exactly one BPS soliton connecting each pair of distinct vacua \cite{Cecotti:1992rm}. As reviewed in Appendix \ref{app:LG}, the BPS equation is a gradient-flow equation determined by the K\"ahler metric, and the image of any BPS soliton under $W$ is the straight segment joining its two critical values. The soliton index may therefore be computed as the intersection number of the corresponding vanishing cycles transported to a common regular fiber.

Over the regular value $w_\ast=0$ the fiber consists of the $m+1$ points $X=0$ and $X=R\,e^{2\pi i j/m}$, $j=0,\ldots,m-1$, with $R=(m+1)^{1/m}$ and $m=k+1$. Transporting along the radial segments $[0,w_\alpha]$, the colliding pair for $\alpha=0$ is found on the real axis, where $W$ maps $[0,1]$ and $[1,R]$ monotonically onto $[w_0,0]$: the sheets through $X=0$ and $X=R$ merge at $X_0=1$.  Since $W\!\left(e^{2\pi i/m}X\right)=e^{2\pi i/m}W(X)$, the sheet through $X=0$ participates in every collision, and the vanishing cycles form a star,
\begin{equation}
    \Delta_\alpha=e_0-e_{\alpha+1}, \qquad \Delta_\alpha\circ\Delta_\beta=1+\delta_{\alpha \beta}, \label{eq:ATypeStarCycles}
\end{equation}
where $e_0$ is the zero-cycle supported at $X=0$, $e_{\alpha+1}$ the one at $R\,e^{2\pi i \alpha/m}$, and $e_i\circ e_j=\delta_{ij}$.  Because the critical values lie on a circle, the segment $[w_\alpha,w_\beta]$ is homotopic to $[w_\alpha,0]\cup[0,w_\beta]$ without crossing further critical values. The soliton index is then the intersection number in this basis:
\begin{equation}
    |N_{\alpha \beta}| = |\Delta_\alpha\circ\Delta_\beta| = 1, \qquad \alpha\neq \beta. \label{eq:ATypeSolitonCount}
\end{equation}
With these orientations every signed multiplicity is $+1$, and ordering the vacua as $\alpha=0,\ldots,k$, the upper-triangular soliton matrix is
\begin{equation}
    N_{\alpha \beta} =
    \begin{cases}
        1, &\alpha<\beta,\\[1mm]
        0, &\alpha\geq \beta,
    \end{cases}
    \qquad \alpha,\beta=0,\ldots,k.
\end{equation}
In matrix form we have
\begin{equation}
    N =
    \begin{pmatrix}
        0&1&1&\cdots&1\\
        0&0&1&\cdots&1\\
        0&0&0&\ddots&\vdots\\
        \vdots&\vdots&\ddots&\ddots&1\\
        0&0&\cdots&0&0
    \end{pmatrix}.
\end{equation}
The associated Stokes matrix is
\begin{equation}
    S = \mathbf 1+N =
    \begin{pmatrix}
        1&1&1&\cdots&1\\
        0&1&1&\cdots&1\\
        0&0&1&\ddots&\vdots\\
        \vdots&\vdots&\ddots&\ddots&1\\
        0&0&\cdots&0&1
    \end{pmatrix}.
    \label{eq:AMinimalStokes}
\end{equation}
The screening pairing is obtained by symmetrizing the Stokes matrix:\footnote{For an LG superpotential with $n$ chiral fields, the vanishing cycles have real dimension $n-1$. Their intersection form obeys $\Delta_\alpha\circ\Delta_\beta=(-1)^{n-1}\Delta_\beta\circ\Delta_\alpha$.
In the present example there is one chiral field, so the vanishing cycles are zero-dimensional and the pairing is symmetric: $\mu=S+S^T$.}
\begin{equation}
    \mu = S+S^T =
    \begin{pmatrix}
        2&1&1&\cdots&1\\
        1&2&1&\cdots&1\\
        1&1&2&\ddots&\vdots\\
        \vdots&\vdots&\ddots&\ddots&1\\
        1&1&\cdots&1&2
    \end{pmatrix}.
    \label{eq:ATypeMuMAtrix}
\end{equation}

Although \eqref{eq:ATypeMuMAtrix} is not written in the usual simple-root basis, it is equivalent to the finite $A$-type Cartan matrix.
To see this, introduce the unimodular matrix
\begin{equation}
    P =
    \begin{pmatrix}
        1&0&0&\cdots&0\\
        -1&1&0&\cdots&0\\
        0&-1&1&\ddots&\vdots\\
        \vdots&\ddots&\ddots&\ddots&0\\
        0&\cdots&0&-1&1
    \end{pmatrix},
    \qquad \det P=1.
\end{equation}
A direct computation gives
\begin{equation}
    P^T \mu P =
    \begin{pmatrix}
        2&-1&0&\cdots&0\\
        -1&2&-1&\cdots&0\\
        0&-1&2&\ddots&\vdots\\
        \vdots&\vdots&\ddots&\ddots&-1\\
        0&0&\cdots&-1&2
    \end{pmatrix}
    = C_{A_m},
    \label{eq:ATypeCartan}
\end{equation}
where $C_{A_m}$ is the $A$-type Cartan matrix.

Since multiplication on the left and right by unimodular matrices does not change the cokernel,
\begin{equation}
    \operatorname{coker}(\mu) \cong \operatorname{coker}\left(C_{A_m}\right).
\end{equation}
The Smith normal form is
\begin{equation}
    \operatorname{SNF}(\mu) = \operatorname{diag} \left( \underbrace{1,\ldots,1}_{m-1}, m+1 \right).
\end{equation}
Using $m=k+1$, it follows that
\begin{equation}
    \mathbb{B}=\operatorname{coker}(\mu) \cong \mathbb Z_{k+2}.
\end{equation}
The matrix $\mu$ is nondegenerate, so there is no free component. This example demonstrates that the BSSG can be extracted entirely from the massive deformation and its soliton data, without appealing to a geometric target space.

\subsubsection{Adding More Massive Fields}

We can also consider the closely related family of Landau-Ginzburg models obtained by including additional fields with a mass term,\footnote{For $I=2$, the zero set of the superpotential $W_0= \frac{1}{k+2}X^{k+2} + Y_1^2+Y_2^2$ is the standard $A$-type hypersurface singularity.}
\begin{equation}
    W_0(X,Y_i) = \frac{1}{k+2}X^{k+2} + \sum_{i=1}^{I}Y_i^2,
\end{equation}
together with the deformation
\begin{equation}
    W(X,Y_i) = \frac{1}{k+2}X^{k+2} -X + \sum_{i=1}^{I}Y_i^2.
\end{equation}
The additional quadratic directions do not modify the vacuum or soliton data. Indeed, the critical-point equations are
\begin{equation}
    X^{k+1}=1, \qquad Y_i=0,
\end{equation}
so each quadratic field has a unique critical point and introduces no additional massive vacua. Moreover, the BPS equation in each $Y_i$ direction is proportional to $Y_i$ itself. Since every soliton approaches $Y_i=0$ at both asymptotic vacua, the corresponding finite-energy solution has $Y_i=0$ throughout the flow. Thus all BPS solitons lie entirely in the $X$-sector, and the soliton matrix $N$ and Stokes matrix $S$ are precisely those of the one-field model discussed above.

What does change is the intersection pairing of the vanishing cycles. For a superpotential in $n$ chiral fields, the vanishing cycles have real dimension $n-1$, and their intersection form satisfies
\begin{equation}
    \Delta_\alpha\circ\Delta_\beta = (-1)^{n-1}\Delta_\beta\circ\Delta_\alpha = (-1)^{I}\Delta_\beta\circ\Delta_\alpha,
\end{equation}
since $n=I+1$. Consequently, the screening pairing associated to the same Stokes matrix is
\begin{equation}
    \mu_I =
    \begin{cases}
        S+S^T, & I \text{ even},\\[2mm]
        S-S^T, & I \text{ odd}.
    \end{cases}
\end{equation}
For $I$ even, this is the symmetric pairing studied above.
For $I$ odd, the pairing is anti-symmetric. Its cokernel is
\begin{equation}
    \mathbb{B}=\operatorname{coker}(\mu_I) \cong
    \begin{cases}
        0, & k\ \text{odd},\\[1mm]
        \mathbb Z, & k\ \text{even},
    \end{cases}
\end{equation}
and in either case it contains no torsion.

\subsection{\texorpdfstring{$\mathcal{D}$}{D}-type Minimal Models} \label{app:DTypeMinimalModel}

The $\mathcal{D}_r$ minimal model is the infrared fixed point of the Landau-Ginzburg theory
\begin{equation}
    W_0(X,Y) = \frac{1}{r-1}X^{r-1}+XY^2.
\end{equation}
This corresponds to the level-$k$ $\mathcal{D}$-type minimal model with
\begin{equation}
    r=\frac{k}{2}+2,
\end{equation}
where $k$ is even. The superpotential has a single degenerate critical point at $X=Y=0$. In order to resolve
it into isolated massive vacua and their solitons, we introduce the relevant deformation
\begin{equation}
    W(X,Y) = \frac{1}{r-1}X^{r-1}+XY^2-X-\varepsilon Y.
    \label{eq:DeformedDType}
\end{equation}

The critical-point equations are
\begin{equation}
    \frac{\partial W}{\partial X} = X^{r-2}+Y^2-1 = 0, \qquad \frac{\partial W}{\partial Y} = 2XY-\varepsilon = 0. \label{eq:DtypeCriticalPoints}
\end{equation}
The second equation gives
\begin{equation}
    Y=\frac{\varepsilon}{2X}.
\end{equation}
Substituting this into \eqref{eq:DtypeCriticalPoints} gives
\begin{equation}
    X^r-X^2+\frac{\varepsilon^2}{4}=0.
    \label{eq:DTypeVacuumPolynomial}
\end{equation}
For generic $\varepsilon$, this degree-$r$ polynomial has $r$ distinct roots. Consequently, the massive theory has $r$ isolated vacua,
\begin{equation}
    (X_\alpha,Y_\alpha) = \left( X_\alpha,\frac{\varepsilon}{2X_\alpha}\right), \qquad \alpha=1,\ldots,r,
\end{equation}
where the $X_\alpha$ are the roots of \eqref{eq:DTypeVacuumPolynomial}. At a critical point, the relation
\begin{equation}
    \frac{\varepsilon^2}{4} = X_\alpha^2-X_\alpha^r
\end{equation}
may be used to write the corresponding critical value as
\begin{equation}
    w_\alpha = W(X_\alpha,Y_\alpha) = \frac{r}{r-1}X_\alpha^{r-1}-2X_\alpha. \label{eq:DTypeCriticalValues}
\end{equation}
For generic $\varepsilon$, these critical values are distinct.

The structure of the vacua is particularly transparent for small $\varepsilon$. In the limit $\varepsilon\to0$, $r-2$ vacua approach
\begin{equation}
    (X_\alpha,Y_\alpha) = \left(\exp\left(\frac{2\pi i \alpha}{r-2}\right), 0 \right), \qquad \alpha=0,\ldots,r-3,
\end{equation}
while the remaining two approach
\begin{equation}
    (X,Y)=(0,1), \qquad (X,Y)=(0,-1).
\end{equation}
The corresponding $r-2$ outer critical values approach equally spaced points on a circle, while the remaining two approach the origin from opposite directions.

The configuration of critical values does not by itself select a unique distinguished basis of vanishing paths. Different choices of distinguished paths are related by Picard--Lefschetz transformations. For the $\mathcal{D}_r$ minimal model, one may choose a distinguished basis in which the nonzero elementary soliton indices form an oriented $D_r$ Dynkin graph \cite{Cecotti:1992rm}. We choose the orientation
\begin{equation}
    1\longrightarrow2\longrightarrow\cdots\longrightarrow r-2, \qquad r-2\longrightarrow r-1, \qquad r-2\longrightarrow r. \label{eq:DTypeOrientation}
\end{equation}
The upper-triangular soliton matrix is therefore
\begin{equation}
    N_{\alpha \beta}
    =
    \begin{cases}
        1, & \beta=\alpha+1,\quad \alpha=1,\ldots,r-3,\\[1mm]
        1, & (\alpha,\beta)=(r-2,r-1),\\[1mm]
        1, & (\alpha,\beta)=(r-2,r),\\[1mm]
        0, & \text{otherwise}.
    \end{cases}
    \label{eq:DTypeSolitonEntries}
\end{equation}
The associated Stokes matrix is
\begin{equation}
    S=\mathbf 1+N.
    \label{eq:DMinimalStokes}
\end{equation}
In matrix form,
\begin{equation}
    S=
    \begin{pmatrix}
        1&1&0&\cdots&0&0&0\\
        0&1&1&\cdots&0&0&0\\
        0&0&1&\ddots&0&0&0\\
        \vdots&\vdots&\ddots&\ddots&1&0&0\\
        0&0&0&\cdots&1&1&1\\
        0&0&0&\cdots&0&1&0\\
        0&0&0&\cdots&0&0&1
    \end{pmatrix}.
    \label{eq:DTypeStokesMatrix}
\end{equation}

This Landau-Ginzburg model contains $n=2$ chiral fields, so the vanishing cycles have real dimension $n-1=1$. Their intersection pairing is therefore anti-symmetric,
\begin{equation}
    \Delta_\alpha\circ\Delta_\beta = -\Delta_\beta\circ\Delta_\alpha.
\end{equation}
The screening pairing is consequently
\begin{equation}
    \mu = S-S^T =
    \begin{pmatrix}
        0&1&0&\cdots&0&0&0\\
        -1&0&1&\cdots&0&0&0\\
        0&-1&0&\ddots&0&0&0\\
        \vdots&\vdots&\ddots&\ddots&1&0&0\\
        0&0&0&\cdots&0&1&1\\
        0&0&0&\cdots&-1&0&0\\
        0&0&0&\cdots&-1&0&0
    \end{pmatrix}.
    \label{eq:DTypeMuMatrix}
\end{equation}

The Smith normal form of $\mu$ depends on the parity of $r$. For odd $r$,
\begin{equation}
    \operatorname{SNF}(\mu) = \operatorname{diag} \left( \underbrace{1,\ldots,1}_{r-1}, 0 \right),
\end{equation}
whereas for even $r$,
\begin{equation}
    \operatorname{SNF}(\mu) = \operatorname{diag} \left( \underbrace{1,\ldots,1}_{r-2}, 0,0 \right).
\end{equation}
It follows that
\begin{equation}
    \mathbb B = \operatorname{coker}(\mu)\cong
    \begin{cases}
        \mathbb Z, & r\ \mathrm{odd},\\[1mm]
        \mathbb Z^2, & r\ \mathrm{even}.
    \end{cases}
\end{equation}
In particular, the cokernel contains no torsion.

A final comment is that the pairing here is anti-symmetric, whereas in the $\mathcal{A}$-type model we had a symmetric pairing. This has to do with the number of fields participating in the definition of the model. Observe that we can also consider a $\mathbb{Z}_2$ gauging of an $\mathcal{A}$-type minimal model and thus reach a $\mathcal{D}$-type minimal model. This changes the statistics of the solitons to an anti-symmetric pairing (see footnote \ref{foot:STAT}). The connection between the two sorts of ``minimal models'' is also apparent once we require the same number of massive fields to be present.

\subsubsection{Adding More Massive Fields}

As in the $\mathcal{A}$-type case, we can consider a closely related family of Landau-Ginzburg models obtained by adding quadratic, or mass, terms to the $\mathcal{D}$-type minimal model\footnote{For $I=1$, the superpotential $W_0=\frac{1}{r-1}X^{r-1} +XY^2 + Z_1^2$ is the standard $D$-type hypersurface singularity.}
\begin{equation}
    W_0(X,Y,Z_i) = \frac{1}{r-1}X^{r-1} +XY^2 +\sum_{i=1}^{I} Z_i^2,
\end{equation}
together with the deformation
\begin{equation}
    W(X,Y,Z_i) = \frac{1}{r-1}X^{r-1} +XY^2 -X -\varepsilon Y +\sum_{i=1}^{I} Z_i^2.
\end{equation}
As before, the additional quadratic fields do not modify the vacuum or soliton data. Indeed, their critical-point equations are simply
\begin{equation}
    Z_i=0,
\end{equation}
so the critical points of the full theory are in one-to-one correspondence with those of the two-field model discussed above. Likewise, the soliton matrix $N$ and Stokes matrix $S$ are unchanged.

The total number of chiral fields is now
\begin{equation}
    n=I+2.
\end{equation}
The vanishing cycles therefore have real dimension
\begin{equation}
    n-1=I+1,
\end{equation}
and their intersection form satisfies
\begin{equation}
    \Delta_\alpha\circ\Delta_\beta = (-1)^{n-1}\Delta_\beta\circ\Delta_\alpha = (-1)^{I+1}\Delta_\beta\circ\Delta_\alpha.
\end{equation}
Consequently, the screening pairing associated to the same Stokes matrix is
\begin{equation}
    \mu_I =
    \begin{cases}
        S-S^T, & I\ \mathrm{even},\\[2mm]
        S+S^T, & I\ \mathrm{odd}.
    \end{cases}
\end{equation}

For $I$ even, the pairing is anti-symmetric and coincides with the pairing of the two-field model discussed above:
\begin{equation}
    \mathbb B = \operatorname{coker}(\mu_I) \cong
    \begin{cases}
        \mathbb Z, & r\ \mathrm{odd},\\[1mm]
        \mathbb Z^2, & r\ \mathrm{even}.
    \end{cases}
\end{equation}
For $I$ odd, the pairing is symmetric,
\begin{equation}
    \mu_I=S+S^T.
\end{equation}
Up to an integral change of basis, this is the finite $D_r$ Cartan matrix. Its cokernel is
\begin{equation}
    \mathbb B = \operatorname{coker}(\mu_I) \cong
    \begin{cases}
        \mathbb Z_4, & r\ \mathrm{odd},\\[1mm]
        \mathbb Z_2\oplus\mathbb Z_2, & r\ \mathrm{even},
    \end{cases}
\end{equation}
which is purely torsion.

\subsection{\texorpdfstring{$(A_n,D_m)$}{AD} Argyres-Douglas Theories} \label{sec:AD}

We now consider the Landau-Ginzburg model of the $(A_n,D_m)$ Argyres-Douglas theory \cite{Argyres:1995jj, Shapere:1999xr, Cecotti:2010fi}. The corresponding Calabi-Yau threefold
singularity is specified by the hypersurface $W_0=0$, where
\begin{equation}
  W_0(X,Y,Z,U) = \frac{X^{n+1}}{n+1}+\frac{Y^{m-1}}{m-1}+YZ^2+U^2, \qquad n\geq 1,\quad m\geq 4.
\end{equation}
Importantly, this superpotential is the sum of the superpotentials for the $\mathcal{A}_{n-1}$ and $\mathcal{D}_m$ minimal models considered above. This will allow us to determine the soliton data of the overall theory from the soliton data of the two constituent theories.

As before, we introduce a massive deformation to isolate the degenerate critical point
\begin{equation}
  W=W_0-X-Y-\varepsilon Z.
\end{equation}
The critical-point equations are
\begin{equation}
  X^n=1,\qquad Y^{m-2}+Z^2=1,\qquad 2YZ=\varepsilon,\qquad U=0.
\end{equation}
As in the individual minimal models, the $\mathcal{D}_m$ equations reduce to
\begin{equation}
  Y^m-Y^2+\frac{\varepsilon^2}{4}=0.
\end{equation}
For generic $\varepsilon$, the $\mathcal{A}_{n-1}$ and $\mathcal{D}_m$ factors have $n$ and $m$ isolated vacua respectively. Therefore, the combined theory has a total of $nm$ vacua labeled by the tuple $(\alpha,\beta)$.

We now compute the intersection pairing. Let $S_\mathcal{A}$ and $S_\mathcal{D}$ denote the Stokes matrices of the deformed $\mathcal{A}_{n-1}$ and $\mathcal{D}_m$ models. The Stokes matrix is the tensor product \cite{Cecotti:1992rm, Cecotti:2013lda}
\begin{equation}
  S_{\mathcal{AD}}=S_\mathcal{A}\otimes S_\mathcal{D}.
\end{equation}
In particular, writing $S_\mathcal{A}=\mathbf{1}_n+N_\mathcal{A}$ and $S_\mathcal{D}=\mathbf{1}_m+N_\mathcal{D}$ gives
\begin{equation}
  S_{\mathcal{AD}}=\mathbf{1}_n\otimes \mathbf{1}_m +N_\mathcal{A}\otimes \mathbf{1}_m +\mathbf{1}_n\otimes N_\mathcal{D} +N_\mathcal{A}\otimes N_\mathcal{D}.
\end{equation}
The superpotential contains four chiral fields, so the intersection pairing is antisymmetric:
\begin{equation}
  \mu_{\mathcal{AD}}=S_{\mathcal{AD}}-S_{\mathcal{AD}}^{T} =S_\mathcal{A}\otimes S_\mathcal{D}-S_\mathcal{A}^{T}\otimes S_\mathcal{D}^{T}.
\end{equation}
The BSSG is then given by
\begin{equation}
    \mathbb{B} = \operatorname{coker}(\mu_{\mathcal{AD}}).
\end{equation}

As a concrete example, consider the $(A_2,D_4)$ theory, with superpotential
\begin{equation}
    W =\frac{X^3}{3}+\frac{Y^3}{3} +YZ^2+U^2-X-Y-\varepsilon Z.
\end{equation}
The $\mathcal{A}_1$ sector has two massive vacua, and the $\mathcal{D}_4$ sector has four massive vacua, which gives a total of eight vacua. Using the results of the previous sections, the Stokes matrices of the constituent minimal models are
\begin{equation}
    S_\mathcal{A}=
    \begin{pmatrix}
    1&1\\
    0&1
    \end{pmatrix},
    \qquad
    S_\mathcal{D}=
    \begin{pmatrix}
    1&1&0&0\\
    0&1&1&1\\
    0&0&1&0\\
    0&0&0&1
    \end{pmatrix}.
\end{equation}
Taking their tensor product gives
\begin{equation}
    S_{\mathcal{AD}}=S_\mathcal{A}\otimes S_\mathcal{D} =
    \begin{pmatrix}
    S_\mathcal{D}&S_\mathcal{D}\\
    0&S_\mathcal{D}
    \end{pmatrix}.
\end{equation}
The intersection pairing is therefore
\begin{equation}
    \mu_{\mathcal{AD}}
    =S_{\mathcal{AD}}-S_{\mathcal{AD}}^T =
    \begin{pmatrix}
    S_\mathcal{D}-S_\mathcal{D}^T&S_\mathcal{D}\\
    -S_\mathcal{D}^T&S_\mathcal{D}-S_\mathcal{D}^T
    \end{pmatrix}.
\end{equation}
Explicitly, we obtain
\begin{equation}
    \mu_{\mathcal{AD}}=
    \left(
    \begin{array}{rrrrrrrr}
     0& 1& 0& 0& 1& 1& 0& 0\\
    -1& 0& 1& 1& 0& 1& 1& 1\\
     0&-1& 0& 0& 0& 0& 1& 0\\
     0&-1& 0& 0& 0& 0& 0& 1\\
    -1& 0& 0& 0& 0& 1& 0& 0\\
    -1&-1& 0& 0&-1& 0& 1& 1\\
     0&-1&-1& 0& 0&-1& 0& 0\\
     0&-1& 0&-1& 0&-1& 0& 0
    \end{array}
    \right).
\end{equation}
It follows that the BSSG is
\begin{equation}
    \mathbb{B}_{\mathcal{AD}}=\operatorname{coker}(\mu_{\mathcal{AD}}) \cong\mathbb{Z}_2\oplus\mathbb{Z}_2.
\end{equation}
This reproduces the target space defect group computation of \cite{DelZotto:2020esg} from the LG model perspective.

\subsection{Mirrors of Toric Threefolds} \label{sec:C3OrbifoldsDual}

As another example, consider the local mirror of a toric Calabi-Yau threefold $M$.
M-theory backgrounds of the form $\mathbb{R}^{4,1} \times M$ engineer 5D SCFTs, and their reduction on a circle leads to type IIA backgrounds and corresponding 4D $\mathcal{N}  = 2$ theories.
One can extract the defect group either from a direct computation of the relative homology of the resolved geometry \cite{Albertini:2020mdx, Morrison:2020ool, Tian:2021cif}, or directly from the adjacency matrix of the accompanying BPS quiver for a 4D KK theory \cite{DelZotto:2022fnw}.
Starting from the adjacency matrix $A$ for the accompanying BPS quiver, the anti-symmetrization
yields the intersection form $\mu_{\alpha \beta} = \Delta_{\alpha} \circ \Delta_{\beta} = - \Delta_{\beta} \circ \Delta_{\alpha}$
for the local mirror geometry:
\begin{equation}
\mu = A - A^{T},
\end{equation}
and the defect group follows from $\mathrm{coker}(\mu)$.

This same quantity follows in essentially the same way from the local mirror.
In these toric examples, the mirror is also a local Calabi-Yau, constructed as \cite{Hori:2000kt, Hori:2000ck}:
\begin{equation}
uv = W(x,y) \,\,\, \text{for} \,\,\, x,y\in\mathbb{C}^{\ast} \,\,\, \text{and} \,\,\, u,v \in \mathbb{C}.
\end{equation}
Here, we have:
\begin{equation}
W(x,y) = \underset{a,b}{\sum} w_{ab}x^{a}y^b,
\end{equation}
where the summation over $(a,b)$ runs over the toric grid points. This is characterized by an LG model with twisted
chiral superfields $Y_{i} \sim Y_{i} + 2 \pi$ and corresponding twisted chiral superpotential $W(Y_i)$.
As explained in \cite{Hori:2000ck}, the spectrum of D-branes and intersection pairing
can be extracted directly from the soliton matrix for the LG model.

As a concrete example, consider the orbifold $\mathbb C^3/\mathbb Z_3$ with weights $(1,1,1)$.
A convenient normalized representative of its mirror Landau-Ginzburg superpotential is:
\begin{equation}
    W(x,y) = x+y+\frac{1}{xy} + e^{-t}. \label{eq:C3Z3LGSuperpotential}
\end{equation}
The critical-point equations are
\begin{equation}
    \partial_xW = 1-\frac{1}{x^2y} = 0, \qquad \partial_yW=1-\frac{1}{xy^2}=0.
\end{equation}
They imply
\begin{equation}
    x=y,\qquad x^3=1.
\end{equation}
There are therefore three massive vacua,
\begin{equation}
    v_0=(1,1),\qquad v_1=(\omega,\omega), \qquad v_2=(\omega^2,\omega^2), \qquad \omega=e^{2\pi i/3},
\end{equation}
with critical values\footnote{Note that the constant term $e^{-t}$ can be absorbed by shifting the superpotential.}
\begin{equation}
    W_0=3, \qquad W_1=3\omega, \qquad W_2=3\omega^2.
\end{equation}

In a distinguished chamber, the Stokes matrix is\footnote{The Stokes matrix depends on the chamber, the ordering of the vacua, and the orientations of the thimbles. Changes of chamber act by integral Picard-Lefschetz transformations, $S\mapsto U^{T}SU$ with $U\in GL_N(\mathbb Z)$, and do not change the isomorphism class of the screening quotient.}
\begin{equation}
    S =
    \begin{pmatrix}
        1 & -3 & 3 \\
        0 & 1 & -3 \\
        0 & 0 & 1
    \end{pmatrix}.
    \label{eq:C3Z3Stokes}
\end{equation}
Its anti-symmetrization gives
\begin{equation}
    \mu = S-S^T =
    \begin{pmatrix}
        0 & -3 & 3 \\
        3 & 0 & -3 \\
        -3 & 3 & 0
    \end{pmatrix}.
    \label{eq:C3Z3Pairing}
\end{equation}

The same matrix follows from the orbifold quiver.
With the nodes ordered as $(0,1,2)$ and with $A_{\alpha \beta}$ counting arrows $\alpha \to \beta$, its adjacency matrix is
\begin{equation}
    A  =
    \begin{pmatrix}
        0 & 3 & 0 \\
        0 & 0 & 3 \\
        3 & 0 & 0
    \end{pmatrix},
\end{equation}
and hence
\begin{equation}
    A-A^{T} =
    \begin{pmatrix}
        0 & 3 & -3 \\
        -3 & 0 & 3 \\
        3 & -3 & 0
    \end{pmatrix}
    = -\mu.
\end{equation}

The Smith normal form of $\mu$ is
\begin{equation}
    \operatorname{SNF}(\mu) = \operatorname{diag}(3,3,0).
\end{equation}
It follows that the boundary state screening group is:
\begin{equation}
    \mathbb B = \operatorname{coker}(\mu) \simeq \mathbb Z \oplus \mathbb Z_3 \oplus  \mathbb Z_3. \label{eq:C3Z3ScreeningQuotient}
\end{equation}

This is in accord with the geometric computation of the defect group for the 5D SCFT obtained from M-theory on $\mathbb{C}^{3} / \mathbb{Z}_3$ \cite{Albertini:2020mdx, DelZotto:2022fnw, Tian:2021cif}.

\subsection{Mirror of a Non-Supersymmetric Orbifold}

In the previous example we discussed the LG model associated to the mirror of a toric Calabi-Yau, and
this target space also admits a description as a local Calabi-Yau threefold. More generally, however, the local mirror of a target space
need not be geometric at all. For example, in reference \cite{Vafa:2001ra} mirrors of non-supersymmetric orbifolds are obtained, and the
corresponding LG description is not a conventional geometric phase. Nevertheless, one expects the same
structure of a boundary screening group to apply. The main point is that while the target space breaks supersymmetry, the 2D worldsheet
theory is still supersymmetric; supersymmetry breaking amounts to the insertion of a suitable tachyon operator in the worldsheet dynamics which triggers an RG flow. We have already seen that the boundary state screening group is sensitive to RG flows, so we can consider the structure of $\mathbb{B}$ for the worldsheet CFT both before insertion of the tachyon operator, as well as deep in the IR.

In fact, the target space defect group for precisely this situation was studied in references \cite{Braeger:2024jcj, Braeger:2025rov}.
There is an accompanying non-supersymmetric quiver gauge theory with distinct adjacency matrices for the bosonic and fermionic degrees of freedom. In this setting, the nodes of the quiver still give a basis of fractional branes, and for a non-supersymmetric orbifold $\mathbb{R}^6 / \Gamma$ these are simply the irreducible representations of $\Gamma$. As found in \cite{Braeger:2024jcj, Braeger:2025rov}, the target space defect group
$\mathbb{D} = \mathrm{coker}(\mu_{\mathrm{fermionic}})$, namely, one considers just the fermionic adjacency matrix. Moreover, after the twisted sector tachyon condenses, the structure of the quiver changes (i.e., the singularity is partially resolved) and there is a sequence of transitions in the defect group $\mathbb{D}_{\mathrm{UV}} \rightarrow \cdots \rightarrow \mathbb{D}_{\mathrm{IR}}$.

We now show that the same data can be extracted from the mirror / worldsheet theory. To that end, we now make the mirror Landau-Ginzburg construction to the non-susy orbifold more explicit. We focus on cyclic quotients of the form $\mathbb{R}^6/\mathbb{Z}_N\,(x_1,x_2,x_3,x_4)$. The orbifold quotient defines an action in the $\mathbf{4}$ of $SU(4)$. There is a holomorphic presentation of the orbifold given by defining the weights
\begin{equation}
    z_i=(x_i + x_4)\mod{N}, \qquad i=1,2,3.
\end{equation}
We may then rewrite the orbifold as $\mathbb{C}^3/\mathbb{Z}_N\,(z_1,z_2,z_3)$.

The mirror Landau-Ginzburg model can be obtained in a similar fashion to the supersymmetric case above. A particularly nice shortcut to write down the LG model is given by enumerating the various marginal and relevant operators in the chiral ring of the worldsheet theory. These operators are organized via their worldsheet R-charge:
\begin{equation}
    R_j = \sum_{i=1}^3 \textrm{Frac}\left( \frac{jz_i}{N} \right),
\end{equation}
where $\textrm{Frac}$ denotes the fractional part and $j=1,\dots,N-1$ labels the twisted sectors. An operator with $R_j < 1$ is relevant, one with $R_j=1$ is marginal, and one with $R_j > 1$ is irrelevant (see \cite{Morrison:2004fr, Braeger:2024jcj} for more details). We will only be concerned with the relevant and marginal operators in writing the LG superpotential.\footnote{For the supersymmetric case, there are no relevant twisted-sector operators, only marginal and irrelevant operators.}

To write down the mirror LG model, we introduce three chiral fields $u_i$, $i=1,2,3$, with
\begin{equation}
    W_0=u_1^N+u_2^N+u_3^N.
\end{equation}
The operator in the $j$-th twisted sector maps to the monomial
\begin{equation}
    \mathcal{O}_j \longleftrightarrow \prod_{i=1}^3 u_i^{p_i^{(j)}}, \qquad p_i^{(j)} = N\,\textrm{Frac}\left(\frac{jz_i}{N}\right)  = jz_i\mod{N} .
\end{equation}
In particular,
\begin{equation}
    \sum_{i=1}^3 p_i^{(j)}=N R_j,
\end{equation}
so marginal operators correspond to degree-$N$ monomials, while relevant operators correspond to monomials of degree less than $N$. The mirror superpotential is obtained by adding the corresponding relevant and marginal monomials to $W_0$, subject to the orbifold projection and, for type II strings, the GSO projection.

This is best illustrated by looking at an example. Consider the case of the non-supersymmetric orbifold $\mathbb{C}^3/\mathbb{Z}_5\,(1,1,1)$, which is the holomorphic presentation of $\mathbb{R}^6/\mathbb{Z}_5\,(2,2,2,-1)$. In this case, there is only one relevant operator $\mathcal{O}_1$ in the first twisted sector of the chiral ring and no marginal operators. Its R-charge is
\begin{equation}
    R_1 = \textrm{Frac}\left( \frac{1}{5} \right) +\textrm{Frac}\left( \frac{1}{5} \right) +\textrm{Frac}\left( \frac{1}{5} \right) = \frac{3}{5}<1.
\end{equation}
The corresponding monomial is
\begin{equation}
    \mathcal{O}_1\longleftrightarrow u_1u_2u_3.
\end{equation}
Thus, the mirror LG superpotential is\footnote{The orbifold group of the LG model is $\Gamma=\{(a_1,a_2,a_3)\in\mathbb{Z}_N^3 : \sum_i a_i z_i\equiv 0 \bmod N\}$, acting as $u_i\mapsto e^{2\pi i a_i/N}u_i$. We have that $\Gamma \cong \mathbb{Z}_{N}^2$ for all faithful orbifold actions, i.e., $\textrm{gcd}(z_1,z_2,z_3,N)=1$.}
\begin{equation}
    W = \left\{ u_1^5+u_2^5+u_3^5 + \lambda u_1u_2u_3  \right\}/\!/\left(\mathbb{Z}_5\times\mathbb{Z}_5\right),
\end{equation}
in agreement with the results of \cite{Vafa:2001ra}. Here $\lambda$ is the coupling of the relevant deformation.

We now turn to the intersection pairing of the mirror LG model. As reviewed in section \ref{sec:SCREEN}, the intersection pairing may equivalently be computed as the open-string index between boundary conditions $\alpha$ and $\beta$ \cite{Hori:2000ck, Feng:2002kk, Witten1982ConstraintsOS}:
\begin{equation}
    I_{\alpha \beta}  = \textrm{Tr}_{\mathcal{H}_{\alpha \beta}}(-1)^FF. \label{eq:OrbifoldWittenIndex}
\end{equation}
For the mirror of the orbifold, this index can be evaluated directly. The boundary conditions are labeled by the irreducible representations $\rho_a$ of $\mathbb{Z}_N$, with $a\in\mathbb{Z}_N$. Let
\begin{equation}
    Q=\rho_{z_1}\oplus\rho_{z_2}\oplus\rho_{z_3}
\end{equation}
denote the three-dimensional representation defining the holomorphic action on $\mathbb{C}^3$. The index between boundary conditions $\rho_\alpha$ and $\rho_\beta$ is determined by the exterior powers of $Q$ \cite{Moore:2004yt}:
\begin{equation}
    I_{\alpha \beta} = \sum_{k=0}^{3}(-1)^k \dim \operatorname{Hom}_{\mathbb{Z}_N} \left( \rho_\beta, (\det Q)^{-1/2}\otimes  \Lambda^kQ\otimes\rho_\alpha \right). \label{eq:WittenIndex}
\end{equation}
For later use, these exterior powers are
\begin{equation}
\begin{split}
    \Lambda^0 Q &= \mathbf{1},\\
    \Lambda^1 Q &= \rho_{z_1}\oplus\rho_{z_2}\oplus\rho_{z_3},\\
    \Lambda^2 Q &= \rho_{z_1+z_2}\oplus\rho_{z_1+z_3}\oplus\rho_{z_2+z_3},\\
    \Lambda^3 Q &= \det Q=\rho_{z_1+z_2+z_3}.
\end{split}
\end{equation}
For $Q=\rho_{z_1}\oplus\rho_{z_2}\oplus\rho_{z_3}$, this becomes
\begin{equation}
    I_{\alpha \beta} = \delta_{\beta,\alpha-s} -\sum_{i=1}^3\delta_{\beta,\alpha+z_i-s} +\sum_{i<j}\delta_{\beta,\alpha+z_i+z_j-s} -\delta_{\beta,\alpha+z_1+z_2+z_3-s},\label{eq:ExplicitOrbifoldWittenIndex}
\end{equation}
where $s = (z_1 + z_2 + z_3)/2$.

We now revisit the $\mathbb{C}^3/\mathbb{Z}_5\,(1,1,1)$ example and compute the intersection pairing of the mirror LG model. We do this by directly computing the open-string index between boundary conditions. For the holomorphic orbifold action
\begin{equation}
    Q=\rho_1\oplus\rho_1\oplus\rho_1,
\end{equation}
the index is obtained as
\begin{equation}
    I_{\alpha \beta} = \sum_{k=0}^{3}(-1)^k \dim\operatorname{Hom}_{\mathbb{Z}_5} \left(\rho_\beta,(\det Q)^{-1/2}\otimes \Lambda^kQ\otimes\rho_\alpha \right).
\end{equation}
Writing $(\det Q)^{1/2}=\rho_s$, we have $2s=z_1+z_2+z_3=3\mod{5}$, and hence $s=4\mod{5}$. The index therefore becomes
\begin{equation}
\begin{split}
    I_{\alpha \beta} =&\; \delta_{\beta,\alpha-s} -\sum_{i=1}^3\delta_{\beta,\alpha+z_i-s} +\sum_{i<j}\delta_{\beta,\alpha+z_i+z_j-s} -\delta_{\beta,\alpha+z_1+z_2+z_3-s} \\
    =&\; \delta_{\beta,\alpha+1} -3\delta_{\beta,\alpha+2} +3\delta_{\beta,\alpha+3} -\delta_{\beta,\alpha+4},
\end{split}
\end{equation}
where all labels are understood modulo $5$. Ordering the boundary conditions by $\alpha=0,\ldots,4$, the resulting intersection pairing is
\begin{equation}
    \mu= I_{\alpha \beta} =
    \begin{pmatrix}
        0 & 1 & -3 & 3 & -1 \\
        -1 & 0 & 1 & -3 & 3 \\
        3 & -1 & 0 & 1 & -3 \\
        -3 & 3 & -1 & 0 & 1 \\
        1 & -3 & 3 & -1 & 0
    \end{pmatrix}.
\end{equation}
As expected, the pairing is anti-symmetric. With this in hand, we find that the BSSG is given by
\begin{equation}
    \mathbb{B} = \textrm{coker}(\mu) = \mathbb{Z}_5 \oplus\mathbb{Z}_5 \oplus \mathbb{Z}.
\end{equation}

\section{Soliton Screening and Interfaces} \label{sec:INTERFACE}

Our discussion so far has centered on determining the spectrum of boundary states which cannot be connected by solitons in a single QFT.
Another natural setup is an interface $\mathbf{I}$ between distinct QFTs, i.e., we single out a spatial direction which asymptotes between two QFTs, $\mathcal{T}_1$ and $\mathcal{T}_2$:
\begin{equation}
\mathcal{T}_1 \vert {\mathbf{I}} \vert \mathcal{T}_2,
\end{equation}
in the obvious notation. In this case, the boundary conditions on the two sides need not be related in any way, so it is interesting to ask whether there exist soliton profiles which can pass through the interface. In the specific case of string theory,
such an interface corresponds to a domain wall between distinct target space theories. These can be rather mild, e.g., two different theories
connected by a deformation of a single worldsheet theory, or more non-perturbative, e.g., domain walls which connect distinct perturbative string theories.\footnote{See e.g., \cite{Distler:2009ri, McNamara:2019rup, Heckman:2025wqd, Anastasi:2026cus, Torres:2026vxx}.} An important question in this regard is whether the domain wall in question is stable, or can dissolve / decay away.

Our aim in this section will be to study the spectrum of solitons in the presence of candidate interfaces, and conversely, to determine
properties of soliton transmission through candidate interfaces.

For our purposes, it will suffice to view an interface as a choice of linear map connecting the lattice of boundary states
of the two theories, namely as a map from $\Lambda_{2}$ to $\overline{\Lambda}_{1}$:\footnote{See e.g., \cite{Bachas:2001vj}. For a more sophisticated categorical treatment, see \cite{Gaiotto:2015aoa, Gaiotto:2015zna}.}
\begin{equation}
\mathbf{I} = \underset{\alpha_1 \beta_{2}}{\sum} \mathbf{I}_{\alpha_{1} \beta_{2}} \vert \alpha_1 \rangle \langle \beta_{2} \vert.
\end{equation}
Equivalently, we can view the interface as specifying a boundary state of a larger ``folded'' theory $\mathcal{T}_1 \otimes \overline{\mathcal{T}_2}$, i.e., where we tensor one theory by its CRT conjugate:\footnote{We choose the reflection of CRT in the direction transverse to the interface.}
\begin{equation}
\vert \mathbf{I} \rangle \rangle = \underset{\alpha_1 \overline{\beta}_{2}}{\sum} \mathbf{I}_{\alpha_{1} \overline{\beta}_{2}} \vert \alpha_1 \rangle \vert \overline{\beta_{2}} \rangle.
\end{equation}
From this perspective, the choice of interface is simply specified by a choice of boundary condition, i.e., an element of $\Lambda_{\mathrm{fold}} \equiv \Lambda_1 \otimes \overline{\Lambda}_2$. As such, we can study the totality of possible transmissions by studying solitons connecting possible boundary conditions:
\begin{equation}
\mu_{\mathrm{fold}} : \Lambda_{\mathrm{fold}} \rightarrow \Lambda_{\mathrm{fold}}.
\end{equation}
The corresponding boundary state screening group:
\begin{equation}
\mathbb{B}_{\mathrm{fold}} = \Lambda_{\mathrm{fold}} / \mathrm{im}(\mu_{\mathrm{fold}})
\end{equation}
catalogs the states which cannot be connected by such solitons. In particular, we can view it as specifying equivalence classes of interfaces which obstruct soliton transmission. More precisely, denote two possible interfaces between $\mathcal{T}_1$ and $\mathcal{T}_2$ as $\mathbf{I}$ and $\mathbf{I}^\prime$. These two interfaces are in the same equivalence class if for some integer matrix $K \in M_{n\times m}(\mathbb{Z})$:\footnote{To see this, note that two interfaces represent the same element of $\mathbb{B}_{\mathrm{fold}}$ if their difference is $\mu_{\mathrm{fold}}(k)$ for some $k\in\Lambda_{\mathrm{fold}}$. Choosing integral bases for $\Lambda_1$ and $\overline{\Lambda}_2$, we may write $k=\sum_{\gamma,\delta}K_{\gamma\delta} e_\gamma\otimes\overline{e}_\delta$, where $K\in M_{n\times m}(\mathbb{Z})$. The action of the folded intersection pairing $\mu_{\mathrm{fold}}=\mu_1\otimes\mu_2^T$ on an integer matrix $K$ is
\begin{equation}
\begin{split}
(\mu_{\mathrm{fold}}K)_{\alpha\beta} &=\sum_{\gamma,\delta} (\mu_1)_{\alpha\gamma} (\mu_2^T)_{\beta\delta}K_{\gamma\delta}\\
&=(\mu_1 K\mu_2)_{\alpha\beta}.
\end{split}
\end{equation}
From this we see that two interfaces represent the same element of $\mathbb{B}_{\mathrm{fold}}$ precisely when their difference lies in the image of $\mu_{\mathrm{fold}}$, or equivalently when $\mathbf{I}-\mathbf{I}^\prime=\mu_1 K\mu_2$ for some integral matrix $K$.}
\begin{equation}
    \mathbf{I}_{\alpha\beta} - \mathbf{I}_{\alpha\beta}^\prime = (\mu_1K\mu_2)_{\alpha\beta}.
\end{equation}
A non-zero charge with respect to $\mathbb{B}_{\mathrm{fold}}$ means that the interface cannot fully dissolve / disappear, and moreover, that the spectrum of D-branes / boundary states is partially blocked from transmission across the interface.

A non-trivial consequence of having a conserved charge under $\mathbb{B}_{\mathrm{fold}}$ is that in the target space, the candidate domain wall cannot dissolve / disappear. Of course, this depends to a large extent on the nature of the interface itself. At one extreme, one can consider the completely terminal interface $\mathbf{I} = \emptyset$, i.e., nothing passes through at all. Observe that in the spacetime, this is more akin to having a bubble of nothing separating the two sides of the domain wall, and a priori nothing prevents this bubble from ``eating up'' the entire target space. Perhaps surprisingly, the ``Kronecker interface'' $\mathbf{I}_{\alpha \beta} = \delta_{\alpha \beta}$ is also quite interesting from the perspective of the folded theory.  In particular, the Kronecker interface only identifies corresponding vacua, and because of this, it does not in general represent the trivial element of $\mathbb{B}_{\mathrm{fold}}$. As we shortly explain, the Kronecker interface typically carries a non-trivial charge under $\mathbb{B}_{\mathrm{fold}}$.

Specializing to the case of 2D $\mathcal{N} = (2,2)$ theories, we can extract $\mu_{\mathrm{fold}}$ directly from a signed index, but now where we trace over the larger Hilbert space of states $\mathcal{H}_{1} \otimes \overline{\mathcal{H}}_{2}$ with prescribed boundary conditions $(B_{\alpha} , B_{\overline{\beta}})$ on the left and boundary conditions $(B_{\alpha^{\prime}} , B_{\overline{\beta}^{\prime}})$ on the right.
The resulting index then takes the form:
\begin{equation}
\mu_{\mathrm{fold}} = \mu_{1} \otimes \mu_{2}^{T},
\end{equation}
in the obvious notation. We comment that for geometric target spaces
this is precisely what one would expect from computing the intersection form on $X = X_1 \times X_2$ from that of the $X_i$ and applying a K\"unneth-type formula to the (co)homology.

The rest of this section is organized as follows. We begin by showing how $\mathbb{B}_{\mathrm{fold}}$ is related to the boundary state screening groups $\mathbb{B}_1$ and $\mathbb{B}_2$. After this, we study some illustrative examples of interfaces between theories. To begin, we consider an interface in which we have a pair of LG theories connected by a deformation which is activated on one side. In this case, the vacua and also spectrum of solitons can be quite different on the two sides of the interface, and we determine candidate charges for interfaces between the two theories. As another example, we study the interface between two $\mathcal{A}$-type minimal models. Appendix \ref{app:LGinterface} discusses some additional aspects of solitons in the presence of an interface.

\subsection{Relating \texorpdfstring{$\mathbb{B}_1$}{B1}, \texorpdfstring{$\mathbb{B}_2$}{B2}, and \texorpdfstring{$\mathbb{B}_{\mathrm{fold}}$}{Bfold}} \label{ssec:BfoldComp}

Before moving on to studying explicit examples, we remark that $\mathbb{B}_{\textrm{fold}}$ can be related directly to $\mathbb{B}_1$ and $\mathbb{B}_2$, i.e., the BSSG of $\mathcal{T}_1$ and $\mathcal{T}_2$ respectively. To see this, let $\Lambda_1,\Lambda_2$ be free abelian lattices. Then $\Lambda_{\textrm{fold}} = \Lambda_1 \otimes \overline{\Lambda}_2$ and
\begin{equation}
    \mathbb{B}_{\textrm{fold}} = \frac{\Lambda_1 \otimes \overline{\Lambda}_2}{\operatorname{im}(\mu_1)\otimes\operatorname{im}(\mu_2^T)}.
\end{equation}
Now introduce the following as an intermediate quotient
\begin{equation}
    \frac{\Lambda_1 \otimes \overline{\Lambda}_2}{\operatorname{im}(\mu_1)\otimes\overline{\Lambda}_2} \cong \mathbb{B}_1 \otimes \overline{\Lambda}_2.
\end{equation}
The kernel of the natural projection onto this quotient is\footnote{Note that $\mu_2$ and $\mu_2^T$ have the same Smith normal form.}
\begin{equation}
    \frac{\operatorname{im}(\mu_1) \otimes \overline{\Lambda}_2}{\operatorname{im}(\mu_1)\otimes\operatorname{im}(\mu_2^T)} \cong \operatorname{im}(\mu_1)\otimes \mathbb{B}_2.
\end{equation}
This gives the following exact sequence:
\begin{equation}
0\longrightarrow \operatorname{im}(\mu_1)\otimes \mathbb{B}_2 \longrightarrow \mathbb{B}_{\mathrm{fold}} \longrightarrow \mathbb{B}_1\otimes\overline{\Lambda}_2 \longrightarrow 0.
\end{equation}
Importantly, this sequence need not split, so $\mathbb{B}_{\mathrm{fold}}$ can contain nontrivial extensions of the screening groups associated with the individual theories.

To make these considerations more concrete, consider the Smith normal form (SNF) of $\mu_1$ and $\mu_2$:
\begin{align}
\mathrm{SNF}(\mu_1) &= \mathrm{diag}(\lambda^{(1)}_{1},\ldots,\lambda^{(1)}_{M_1}), \\
\mathrm{SNF}(\mu_2) &= \mathrm{diag}(\lambda^{(2)}_{1},\ldots,\lambda^{(2)}_{M_2}),
\end{align}
where the entries are all non-negative integers. From this we conclude that the boundary state screening group
for the two theories is:
\begin{align}
\mathbb{B}_1 = \underset{i}{\bigoplus} \mathbb{Z} / \lambda^{(1)}_{i} \mathbb{Z}, \\
\mathbb{B}_2 = \underset{j}{\bigoplus} \mathbb{Z} / \lambda^{(2)}_{j} \mathbb{Z}.
\end{align}
Turning next to the image of $\mu_1 \otimes \mu_{2}^{T}$, we conclude that the boundary state screening group of the folded theory is therefore:
\begin{equation}
\mathbb{B}_{\mathrm{fold}} = \underset{i,j}{\bigoplus} \mathbb{Z} / \lambda^{(1)}_{i} \lambda^{(2)}_{j} \mathbb{Z},
\end{equation}
in line with the more abstract considerations presented above. In particular, (for non-zero Smith invariants) we can construct short exact sequences:
\begin{align}
0 &\longrightarrow \mathbb{Z} / \lambda^{(1)}_{i} \mathbb{Z} \longrightarrow \mathbb{Z} / \lambda^{(1)}_{i} \lambda^{(2)}_{j} \mathbb{Z} \longrightarrow \mathbb{Z} / \lambda^{(2)}_{j} \mathbb{Z} \longrightarrow 0, \\
0 &\longrightarrow \mathbb{Z} / \lambda^{(2)}_{j} \mathbb{Z} \longrightarrow \mathbb{Z} / \lambda^{(1)}_{i} \lambda^{(2)}_{j} \mathbb{Z} \longrightarrow \mathbb{Z} / \lambda^{(1)}_{i} \mathbb{Z} \longrightarrow 0,
\end{align}
so taking direct sums over the $i$ and $j$ indices yields:
\begin{align}
0 &\longrightarrow \mathbb{B}_{1}^{r_2} \longrightarrow \mathbb{B}_{\mathrm{fold}} \longrightarrow \mathbb{B}_{2}^{M_1} \longrightarrow 0, \\
0 &\longrightarrow \mathbb{B}_{2}^{r_1} \longrightarrow \mathbb{B}_{\mathrm{fold}} \longrightarrow \mathbb{B}_{1}^{M_2} \longrightarrow 0,
\end{align}
where $M_i = \textrm{rank}(\Lambda_i)$ and $r_i = \textrm{rank}(\mu_i)$.

We now illustrate some of these general considerations by way of example.

\subsection{Geometric Example: SymTrees and Worldsheet Interfaces}

As a first class of examples, consider type IIA string theory on the $A$-type singularity $\mathbb{C}^2 / \mathbb{Z}_{K}$. This yields 6D $\mathcal{N} = (1,1)$ Super Yang-Mills theory with gauge algebra $\mathfrak{su}(K)$, with the global form of the gauge group dictated by the choice of boundary conditions for the RR potentials of the system. The orbifold can be presented as the hypersurface singularity:
\begin{equation}
uv = z^{K}.
\end{equation}
The defect group for the corresponding background is $\mathbb{D} = \mathbb{Z}_{K}^{(1)} \oplus \mathbb{Z}_{K}^{(3)}$ and a choice of polarization dictates the global form of the symmetry.

Suppose we now perform a deformation of the original singularity to:
\begin{equation}
uv = (z-z_1)^{K_1} (z-z_2)^{K_2},
\end{equation}
with $K = K_1 + K_2$. Near $z = z_{i}$ we have an isolated singularity with corresponding defect group $\mathbb{D}_{i}$, but far away, the boundary geometry is still an $S^{3} / \mathbb{Z}_{K}$. The jump in the defect group is captured in terms of an accompanying ``SymTree'' structure which amounts to a fusing of different SymTFTs at a junction theory \cite{Baume:2023kkf}.

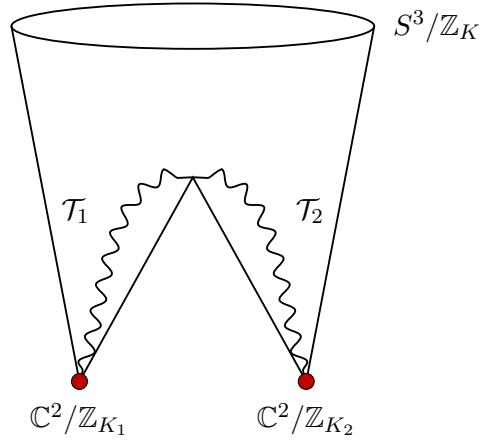
\begin{figure}
\centering
\begin{tikzpicture}[font=\small]
  \coordinate (L) at (-1.5,0);
  \coordinate (R) at ( 1.5,0);
  \coordinate (C) at ( 0,2.7);

  \draw[ThickLine] (-2.4,4.7) -- (L) -- (C) -- (R) -- (2.4,4.7);

  \draw[ThickLine] (0,4.7) ellipse (2.4 and 0.3);
  \node[right=3pt] at (2.4,4.7) {$S^3/\mathbb{Z}_K$};

  \draw[ThickLine, snake it, decoration={pre length=4pt, post length=1pt}]
    (L) to[out=72,in=180] (C);
  \draw[ThickLine, snake it, decoration={mirror, pre length=4pt, post length=1pt}]
    (R) to[out=108,in=0] (C);
  \node at (-1.55,2.25) {$\mathcal{T}_1$};
  \node at ( 1.55,2.25) {$\mathcal{T}_2$};

  \node[SmallCircleRed] at (L) {};
  \node[SmallCircleRed] at (R) {};
  \node[below=4pt] at (L) {$\mathbb{C}^2/\mathbb{Z}_{K_1}$};
  \node[below=4pt] at (R) {$\mathbb{C}^2/\mathbb{Z}_{K_2}$};
\end{tikzpicture}
\caption{A long string stretched between the two singularities. Its worldsheet is described by $\mathcal{T}_1$ near the first singularity and by $\mathcal{T}_2$ near the second singularity. The point where the string crosses the junction traces out a line on the worldsheet. This line is the interface between $\mathcal{T}_1$ and $\mathcal{T}_2$.}
\label{fig:INTERFACETARGETSPACE}
\end{figure}

Let us now turn to the accompanying worldsheet characterization of the defect group. To this end, consider a long string which stretches from the $z = z_1$ singularity to the $z = z_2$ singularity. Such a string can attach to D-branes touching the tips of the two singularities. These branes correspond to choices of boundary conditions in two distinct theories, namely $\mathcal{T}_1$ and $\mathcal{T}_2$. The junction of the SymTree in this example corresponds to the data of the interface. Observe that some D-branes can stretch between the two singularities. These are the massive W-bosons from the breaking pattern $\mathfrak{su}(K) \supset \mathfrak{su}(K_1) \times \mathfrak{su}(K_2)\times \mathfrak{u}(1)$. In the LG description, these correspond to solitonic configurations connecting two vacua across an interface. In the target space, this interface corresponds to the intersection locus between the long string and the junction theory used to build the SymTree (see figure \ref{fig:INTERFACETARGETSPACE}).

Let us now turn to the LG description of the same system. It is enough to consider a parent theory:
\begin{equation}
W_{\mathrm{parent}} = UV - Z^{K}.
\end{equation}
The boundary screening group for this system is $\mathbb{Z}_K$. Next, consider two different relevant deformations on the left and right of the 2D worldsheet, which we scale so as to eliminate some of the massive vacua of the original theory.
On the left, we deform to $W_1 = UV - Z^{K_1}$, while on the right we deform to $W_2 = UV - Z^{K_2}$.
In this setting, the interface corresponds to the intermediate region where the deformation is still finite.

Similar considerations hold for related systems. In particular, most of the SymTree examples engineered in reference \cite{Baume:2023kkf}
have direct LG model avatars.

\subsection{Illustrative LG Examples}

In this section we consider various LG examples of interfaces between distinct $\mathcal{A}$-type minimal models. In particular, we will focus on the case where $\mathcal{T}_1$ and $\mathcal{T}_2$ are both $\mathcal{A}_2$ minimal models with superpotentials
\begin{equation}
    W_i = \frac{1}{4}X_i^4,
\end{equation}
where $i=1,2$ refers to the two theories. We consider a variety of interfaces $\mathbf{I}$ that illustrate the general points made above.

In this case the intersection pairings are
\begin{equation}
    \mu_1 = \mu_2 =
    \begin{pmatrix}
        2 & -1 & 0 \\
        -1 & 2 & -1 \\
        0 & -1 & 2
    \end{pmatrix}.
\end{equation}
This also gives the intersection pairing in the folded theory:
\begin{equation}
    \mu_\textrm{fold} = \mu_1\otimes\mu_2^T =
    \begin{pmatrix}
        4 & -2 & 0 & -2 & 1 & 0 & 0 & 0 & 0 \\
        -2 & 4 & -2 & 1 & -2 & 1 & 0 & 0 & 0 \\
        0 & -2 & 4 & 0 & 1 & -2 & 0 & 0 & 0 \\
        -2 & 1 & 0 & 4 & -2 & 0 & -2 & 1 & 0 \\
        1 & -2 & 1 & -2 & 4 & -2 & 1 & -2 & 1 \\
        0 & 1 & -2 & 0 & -2 & 4 & 0 & 1 & -2 \\
        0 & 0 & 0 & -2 & 1 & 0 & 4 & -2 & 0 \\
        0 & 0 & 0 & 1 & -2 & 1 & -2 & 4 & -2 \\
        0 & 0 & 0 & 0 & 1 & -2 & 0 & -2 & 4
\end{pmatrix}.
\end{equation}
From this we compute
\begin{equation}
    \mathbb{B}_{\textrm{fold}} = \operatorname{coker}(\mu_\textrm{fold}) = \mathbb{Z}_4^4 \oplus \mathbb{Z}_{16}.
\end{equation}

We first consider the Kronecker interface between the two $\mathcal{A}_2$ minimal models. This interface acts as the identity map between the Hilbert spaces of the two theories: $\mathbf{I}_{\textrm{Kron}}=\sum_{\alpha=1}^{3}|\alpha_1\rangle\langle\alpha_2|$. Written as a matrix, we have
\begin{equation}
\mathbf{I}_\textrm{Kron}=
\begin{pmatrix}
1&0&0\\
0&1&0\\
0&0&1
\end{pmatrix}.
\end{equation}
Although the two theories are identical, the interface only identifies corresponding vacua, so off-diagonal connections between distinct vacua on opposite sides are absent. Thus, as expected, after folding this theory carries a nontrivial charge under $\mathbb{B}_{\mathrm{fold}}=\mathbb{Z}_4^4\oplus\mathbb{Z}_{16}$. In particular, it carries a charge $(2,3,2,3,14) \in \mathbb{Z}_4^4\oplus \mathbb{Z}_{16}$. In order to determine this charge, we first choose a basis for $\mu_1=\mu_2=\mu$, with $U\mu V=D$ where $D=\operatorname{diag}(1,1,4)$ and where $U,V$ are unimodular. For example, we may take\footnote{It is important to make the same choice across all examples, as the particular charge under $\mathbb{B}_\textrm{fold}$ is basis dependent. The order of the charge is basis independent.}
\begin{equation}
U=
\begin{pmatrix}
0&1&0\\
0&0&1\\
1&2&3
\end{pmatrix},
\qquad \textrm{and} \qquad
V=
\begin{pmatrix}
-1&-2&3\\
0&-1&2\\
0&0&1
\end{pmatrix}.
\end{equation}
The folded pairing is then diagonalized as $(U\otimes U)\mu_{\textrm{fold}}(V\otimes V) = D\otimes D$, which has four nontrivial entries equal to $4$ and one equal to $16$ as expected. The Kronecker interface transforms as
\begin{equation}
U\mathbf{I}_\textrm{Kron}U^T=
\begin{pmatrix}
1&0&2\\
0&1&3\\
2&3&14
\end{pmatrix}.
\end{equation}
Recall that $D = \operatorname{diag}(1,1,4)$, so we read off the entries $(1,3),(2,3),(3,1),(3,2)$ modulo $4$, and $(3,3)$ modulo $16$. This gives $(2,3,2,3,14)\in\mathbb{Z}_4^4\oplus\mathbb{Z}_{16}$. 

Next, consider the ``reversal'' interface between the two $\mathcal{A}_2$ minimal models. This interface exchanges the first and third vacua while leaving the second vacuum unchanged. Explicitly, it acts as the map
\begin{equation}
\mathbf{I}_{\mathrm{rev}}
=|1_1\rangle\langle 3_2|
+|2_1\rangle\langle 2_2|
+|3_1\rangle\langle 1_2|.
\end{equation}
Written as a matrix, we have
\begin{equation}
\mathbf{I}_{\mathrm{rev}}=
\begin{pmatrix}
0&0&1\\
0&1&0\\
1&0&0
\end{pmatrix}.
\end{equation}
Unlike the Kronecker interface, $\mathbf{I}_{\mathrm{rev}}$ identifies vacua in the opposite ordering, so that the first and third vacua are exchanged.

To determine its charge under $\mathbb{B}_{\mathrm{fold}}$, we use the same change of basis $U$ as in the previous example. Since the interface transforms as $\mathbf{I}_{\mathrm{rev}}\mapsto U\mathbf{I}_{\mathrm{rev}}U^T$, we obtain
\begin{equation}
U\mathbf{I}_{\mathrm{rev}}U^T=
\begin{pmatrix}
1&0&2\\
0&0&1\\
2&1&10
\end{pmatrix}.
\end{equation}
Reading off the entries $(1,3),(2,3),(3,1),(3,2)$ modulo $4$, and $(3,3)$ modulo $16$, gives the charge $(2,1,2,1,10)\in\mathbb{Z}_4^4\oplus\mathbb{Z}_{16}$. This is a nontrivial element of order eight. In particular, it differs from the charge $(2,3,2,3,14)$ carried by the identity interface, and their difference is an element of order four. 

Finally, consider the completely terminal interface $\mathbf{I}=\emptyset$, which blocks all transmissions between the two theories. This corresponds to the zero map between their Hilbert spaces, represented by
\begin{equation}
\mathbf{I}_{\emptyset}=
\begin{pmatrix}
0&0&0\\
0&0&0\\
0&0&0
\end{pmatrix}.
\end{equation}
Unlike the previous examples, no vacua on opposite sides of the interface are identified. As such, this interface carries the trivial charge $(0,0,0,0,0)\in\mathbb{B}_{\mathrm{fold}}$. Indeed, its transformation under the Smith basis is simply $U\mathbf{I}_{\emptyset}U^T=0$. Interestingly, the terminal interface is not the unique representative of its equivalence class. For example, consider the nonzero interface
\begin{equation}
\mathbf{I}^{\prime}=
\begin{pmatrix}
1&0&1\\
0&0&0\\
1&0&1
\end{pmatrix}.
\end{equation}
Although this interface allows additional connections between vacua on opposite sides, it represents the same element of $\mathbb{B}_{\mathrm{fold}}$. To see this, observe that
\begin{equation}
\mathbf{I}^{\prime}-\mathbf{I}_{\emptyset}
=\mu_1
\begin{pmatrix}
1&1&1\\
1&1&1\\
1&1&1
\end{pmatrix}
\mu_2.
\end{equation}
Since the intermediate matrix is integral, the two interfaces belong to the same equivalence class.

\section{Soliton Screening and Junctions} \label{sec:JUNCTION}

\begin{figure}[t!]
\begin{center}
\begin{tikzpicture}[
  line cap=round,
  line join=round,
  sheet/.style={draw=black!65, line width=0.85pt},
  theory/.style={font=\Large, inner sep=2pt}
]

  \definecolor{sheetred}{RGB}{246,214,216}
  \definecolor{sheetblue}{RGB}{214,228,246}

  \coordinate (A) at (0,-1);
  \coordinate (B) at (0, 1);

  \filldraw[sheet, fill=white]
    (A) -- (3.4,1.7)
    .. controls (3.60,2.25) and (3.60,3.15) .. (3.4,3.7)
    -- (B) -- cycle;

  \filldraw[sheet, fill=sheetblue]
    (A) -- (3.4,-3.7)
    .. controls (3.60,-3.15) and (3.60,-2.25) .. (3.4,-1.7)
    -- (B) -- cycle;

  \filldraw[sheet, fill=sheetred]
    (A) -- (-3.8,-1)
    .. controls (-4.0,-0.45) and (-4.0,0.45) .. (-3.8,1)
    -- (B) -- cycle;

  \coordinate (X) at ({3.4/2.7},0);
  \draw[black!55, line width=0.65pt, densely dashed]
    (A) -- (X);

  \draw[black!90, line width=2pt] (A) -- (B);

  \node[theory] at (-2.1, 0)   {$\mathcal{T}_1$};
  \node[theory] at ( 2.2, 1.9) {$\mathcal{T}_2$};
  \node[theory] at ( 2.2,-1.9) {$\mathcal{T}_3$};

  \node[font=\Large, anchor=south, fill=white, inner sep=2pt]
    at (0,1.12) {$\mathbf{J}_{123}$};
\end{tikzpicture}
\end{center}
    \caption{Depiction of QFTs $\mathcal{T}_1$, $\mathcal{T}_{2}$, and $\mathcal{T}_3$ joined by a junction $\mathbf{J}_{123}$. 
    When the junction carries a non-trivial boundary state screening group charge under $\mathbb{B}_{\mathrm{fold}}$, 
    it cannot fully dissolve / disappear.}
    \label{fig:JUNCTION}
\end{figure}
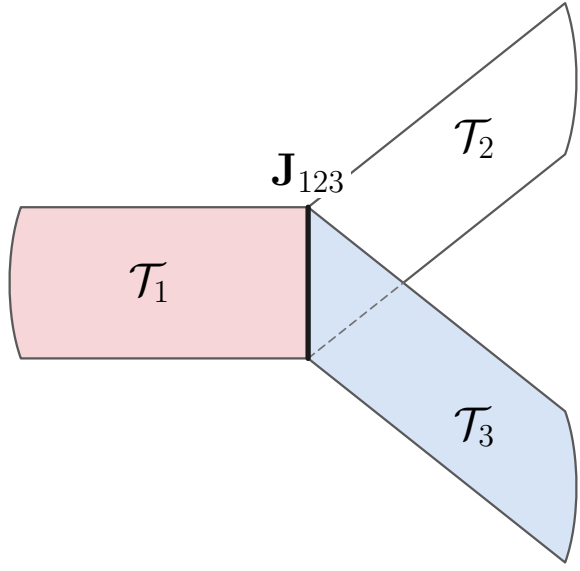

The discussion of the previous section focused on the case of an interface between two QFTs. Similar considerations apply to junctions of QFTs, where we define each QFT $\mathcal{T}_i$ on a half-line $\mathbb{R}_{\leq 0}$, each of which is joined at a common origin (see figure \ref{fig:JUNCTION}). Such configurations are also of interest in the special case of worldsheet CFTs, where the target space interpretation is that of distinct perturbative string theories joined at a non-perturbative interface \cite{Altavista:2026edv, Altavista:2026muc, Tachikawa:2026jaj}. This sort of structure also naturally appears in the SymTree generalization of the SymTFT construction (see e.g., \cite{Baume:2023kkf}).

Let us now discuss in more detail the structure of the boundary state screening group for $N$ QFTs joined by a single junction. To each QFT $\mathcal{T}_i$ we have a $\mathbb{Z}$-module of boundary states $\Lambda_i$, and corresponding index / intersection pairing $\mu_{i}$, and with it a corresponding boundary state screening group $\mathbb{B}_i = \Lambda_i / \mathrm{im}(\mu_i)$. Without loss of generality, we can take the CRT conjugate of all theories so that the junction can be viewed as a boundary state in $\Lambda_{1} \otimes \cdots \otimes \Lambda_{N}$,
i.e., much as in the original ``folding trick'' we can consider the QFT $\mathcal{T}_1 \otimes \cdots \otimes \mathcal{T}_{N}$ on the half-line $\mathbb{R}_{\leq 0}$ with a boundary state:
\begin{equation}
\vert \mathbf{J}_{1\ldots N} \rangle \rangle =\sum_{\alpha^{(1)} \ldots \alpha^{(N)}} \mathbf{J}_{\alpha^{(1)}\ldots\alpha^{(N)}} \vert \alpha^{(1)} \ldots \alpha^{(N)} \rangle,
\end{equation}
where the $\alpha^{(n)}$ range over the basis of boundary states for $\Lambda_{n}$. In this folded theory, we can construct the corresponding
boundary state screening group:
\begin{equation}
\mathbb{B}_{\mathrm{fold}} = \frac{\Lambda_{1} \otimes \cdots \otimes \Lambda_{N}}{\mathrm{im}(\mu_1 \otimes \cdots \otimes \mu_N)}.
\end{equation}
A non-zero charge with respect to $\mathbb{B}_{\mathrm{fold}}$ means that the junction cannot fully dissolve / disappear, and moreover, that the spectrum of D-branes / boundary states is partially blocked from transmission across the junction.

The explicit computation of $\mathbb{B}_{\mathrm{fold}}$ in such a junction of theories proceeds much as in the case of section \ref{ssec:BfoldComp}. Indeed, we can construct the Smith normal form (SNF) of $\mu_n$:
\begin{equation}
\mathrm{SNF}(\mu_n) = \mathrm{diag}(\lambda^{(n)}_{1},\ldots,\lambda^{(n)}_{M_n}),
\end{equation}
where the entries are all non-negative integers. From this we conclude that the boundary state screening group is:
\begin{equation}
\mathbb{B}_n = \underset{i}{\bigoplus} \mathbb{Z} / \lambda^{(n)}_{i} \mathbb{Z}.
\end{equation}
Turning next to the image of $\mu_1 \otimes \cdots \otimes \mu_{N}$, we conclude that
the boundary state screening group of the folded theory is therefore:
\begin{equation}
\mathbb{B}_{\mathrm{fold}} = \underset{i_1,\ldots,i_N}{\bigoplus} \mathbb{Z} / \lambda^{(1)}_{i_1} \cdots \lambda^{(N)}_{i_N} \mathbb{Z}.
\end{equation}

Much as in our previous discussion in section \ref{sec:INTERFACE}, we can construct a range of possible junctions.

\section{Generalizations to Higher Dimensions} \label{sec:HIGHER}

The construction of the boundary state screening group naturally extends to higher-dimensional QFTs. Consider a $D$-dimensional QFT on $\mathbb{R}^{1,D-2}\times I$, with boundary conditions $B_\alpha$ and $B_\beta$ at the two ends of the interval. As before, we are interested in soliton configurations which interpolate between the various different boundary conditions. We assume that the boundary conditions of interest span an integral lattice $\Lambda$, and that these interpolating configurations define an integral screening map $\mu:\Lambda\rightarrow\Lambda$. With this in hand, we can again define the boundary state screening group as
\begin{equation}
    \mathbb{B}=\operatorname{coker}(\mu) =\Lambda/\operatorname{im}(\mu).
\end{equation}
Importantly, in higher dimensions the existence of an integral lattice is not automatic and must be established independently. This construction therefore applies whenever such a suitable integral lattice and screening map can be identified. We now illustrate this in four-dimensional $\mathcal{N}=1$ Wess-Zumino models, where the relevant solitonic configurations are BPS domain walls.

\subsection{4D \texorpdfstring{$\mathcal N=1$}{N=1} Wess-Zumino Models} \label{sec:WZModels}

As a concrete example, we now consider four-dimensional $\mathcal N=1$ Wess-Zumino theories. Consider a theory of chiral multiplets $\Phi^i$ with action
\begin{equation}
    S_{4\mathrm D} = \int d^4x\,d^4\theta\, \mathcal{K}\!\left(\Phi^i,\overline{\Phi}^{\bar\imath}\right) +  \left[ \frac{1}{2} \int d^4x\,d^2\theta\,  W(\Phi^i) + \mathrm{c.c.} \right]. \label{eq:4dWZAction}
\end{equation}
The scalar fields take values in a K\"ahler manifold with metric
\begin{equation}
    g_{i\bar\jmath} = \partial_i\partial_{\bar\jmath}\mathcal{K},
\end{equation}
and the supersymmetric vacua are the critical points of the superpotential,
\begin{equation}
    \partial_iW=0. \label{eq:4dWZVacua}
\end{equation}

Upon dimensional reduction on $T^2$, a four-dimensional $\mathcal N=1$ chiral multiplet becomes a two-dimensional $\mathcal N=(2,2)$ chiral multiplet. Before canonically normalizing the two-dimensional fields, the reduced K\"ahler potential and superpotential are
\begin{equation}
    \mathcal{K}_{2\mathrm D} = \operatorname{Vol}(T^2)\,\mathcal{K}_{4\mathrm D}, \qquad  W_{2\mathrm D} = \operatorname{Vol}(T^2)\,W_{4\mathrm D}. \label{eq:4d2dKWReduction}
\end{equation}
The common volume factor cancels from the gradient-flow equations, so the two theories have the same critical points and the same unparametrized BPS trajectories in field space.

The same correspondence can be seen directly in four dimensions. Consider a static field configuration depending only on one spatial coordinate, which we denote by $x^3$. Such a configuration describes a domain wall extended along the remaining two spatial directions.
The BPS domain-wall equation is given by\footnote{The derivation follows exactly as in the 2D case covered in Appendix \ref{app:LG}.}
\begin{equation}
    \frac{d\phi^i}{dx^3} = \frac{e^{i\alpha}}{2}\, g^{i\bar\jmath} \partial_{\bar\jmath}\overline W, \qquad \alpha = \arg\!\left(W_b-W_a\right), \label{eq:4dWZBPSEquation}
\end{equation}
for a wall interpolating between vacua $v_a$ and $v_b$, and its tension obeys
\begin{equation}
    T_{ab} \geq \left|W_b-W_a\right|, \label{eq:4dWZWallTension}
\end{equation}
with equality for a BPS wall.

Equation~\eqref{eq:4dWZBPSEquation} is the same gradient-flow equation that governs BPS solitons in the corresponding two-dimensional Landau-Ginzburg model. Consequently, the critical points, gradient trajectories, vanishing cycles, and Picard-Lefschetz data associated with $W$ are unchanged. In particular, the same Stokes matrix determines the signed incidence data of the BPS sectors, and the same matrix
\begin{equation}
    \mu = S+\varepsilon\,S^T, \qquad \varepsilon=(-1)^{n-1} \label{eq:4dWZScreeningMatrix}
\end{equation}
determines the corresponding screening relations.

As a simple example, consider the same holomorphic function that appeared in the mirror Landau--Ginzburg description of $\mathbb C^3/\mathbb Z_3(1,1,1)$. Introducing a mass scale $\rho$, a dimensionally consistent four-dimensional superpotential is
\begin{equation}
    W_{4\mathrm d}(X,Y) = \rho^2(X+Y) + \frac{\rho^5}{XY}. \label{eq:4dC3Z3Superpotential}
\end{equation}
In terms of the dimensionless variables
\begin{equation}
    x=\frac{X}{\rho}, \qquad y=\frac{Y}{\rho},
\end{equation}
this becomes
\begin{equation}
    W_{4\mathrm d} = \rho^3 \left( x+y+\frac{1}{xy} \right). \label{eq:4dC3Z3DimensionlessSuperpotential}
\end{equation}
This should be regarded as a four-dimensional effective Wess-Zumino theory with field space $(\mathbb C^\ast)^2$.

The critical-point equations are
\begin{equation}
    \partial_XW_{4\mathrm d} = \rho^2-\frac{\rho^5}{X^2Y} = 0, \qquad \partial_YW_{4\mathrm d} = \rho^2-\frac{\rho^5}{XY^2} = 0.
\end{equation}
They imply
\begin{equation}
    X=Y, \qquad X^3=\rho^3,
\end{equation}
and hence the theory has three supersymmetric vacua,
\begin{equation}
    (X,Y) = (\rho,\rho), \qquad (\rho\omega,\rho\omega), \qquad (\rho\omega^2,\rho\omega^2), \qquad\omega=e^{2\pi i/3}. \label{eq:4dC3Z3Vacua}
\end{equation}
Their critical values are
\begin{equation}
    W_r = 3\rho^3\omega^r, \qquad r=0,1,2. \label{eq:4dC3Z3CriticalValues}
\end{equation}

The BPS domain walls therefore obey exactly the same dimensionless gradient-flow equations as the solitons of section \ref{sec:C3OrbifoldsDual}. Their Stokes and intersection pairing matrices are consequently
\begin{equation}
    S =
    \begin{pmatrix}
        1 & -3 & 3 \\
        0 & 1 & -3 \\
        0 & 0 & 1
    \end{pmatrix},
    \qquad
    \mu=S-S^T=
    \begin{pmatrix}
        0 & -3 & 3 \\
        3 & 0 & -3 \\
        -3 & 3 & 0
    \end{pmatrix}.
    \label{eq:4dC3Z3StokesPairing}
\end{equation}
The non-normalizable asymptotic sectors are graded by the same charge lattice $\Lambda=\bigoplus_r\mathbb Z[\Gamma_r]$, so the four-dimensional screening quotient is
\begin{equation}
    \mathbb B_{4\mathrm D} = \operatorname{coker}(\mu) \simeq \mathbb Z \oplus \mathbb Z_3 \oplus \mathbb Z_3.
\end{equation}

\section{Conclusions} \label{sec:CONC}

Motivated by the structure of the target space defect group in QFTs engineered in string theory, we have introduced the boundary state screening group. This group measures the obstruction to connecting states in the integral lattice of boundary states via formal kink / soliton profiles in massive QFTs. This construction makes sense for QFTs with boundaries, and we have presented several illustrative examples primarily in the case of massive LG models. Interfaces between distinct QFTs can, via the folding trick, be treated as another class of boundary states. In this framework, an interface which carries a non-trivial charge under the boundary state screening group cannot fully dissolve / disappear. Similar considerations apply for junctions of QFTs. In the remainder of this section we discuss some potential
avenues for future investigation.

Interfaces in 2D theories provide a worldsheet analog of spacetime bordisms, with corresponding connections to the theory of topological modular forms (see e.g., \cite{Tachikawa:2025flw}). It would be interesting to investigate potential relations between non-trivial topological modular form classes and the boundary state screening group.

The Swampland cobordism conjecture predicts the existence of domain walls between different perturbative string theories. This includes a predicted domain wall between IIA and IIB string theory \cite{Distler:2009ri, McNamara:2019rup, Heckman:2025wqd, Anastasi:2026cus, Torres:2026vxx}.  From the perspective of a long string which passes through the domain wall, this defines an interface CFT \cite{Heckman:2025wqd}. It would be interesting to extend our analysis of the soliton screening group to this situation as well.

The examples we have presented primarily focused on 2D LG models, though we expect the same framework to work far more generally. A particularly intriguing case to consider is the theory of M2-branes in M-theory probing a geometric target space. In this case, the target space defect group should match to the boundary state screening group for a 3D QFT.

In many cases of interest, the interface theory carries a non-trivial charge, indicating the presence of non-trivial localized modes. It would be interesting to see whether additional protected quantities can be extracted for this and related interfaces.

Finally, we have primarily focused on the case of generic 2D QFTs, but one could also consider the closely related case of holographic CFTs with boundaries. In this case, the boundary state screening group tracks solitonic objects in the bulk AdS dual, carrying a non-zero discrete charge. It would be natural to use this to extract further details on the structure of interfaces, perhaps along the lines of reference \cite{Karch:2022vot}.

\section*{Acknowledgments}

We thank S. Cecotti, M. H\"ubner, S. Meynet, M. Rajaguru, M. Romo, and C. Vafa for helpful discussions.
The work of VC is supported by an NSF Graduate Research Fellowship.
The work of VC and JJH is supported by DOE (HEP) Award DE-SC0013528 and BSF grant 2022100. VC and JJH thank the 2026 Simons Summer workshop for hospitality during part of this work. VC thanks IFT Madrid for hospitality during part of this work.

\section*{AI Usage Attestation}

All of the thinking, writing, and calculations in this work were performed by the authors.
AI was used to cross-check some calculations; to review some aspects of the literature; 
in the compilation of some references; and in the creation of some figures.

\appendix

\section{LG Solitons and D-branes}\label{app:LG}

In this Appendix we briefly review aspects of BPS solitons used in the main text. More details can be found in \cite{Hori:2000ck}. In particular, we explain how the directed soliton matrix $N$ determines the Stokes matrix $S=\mathbf 1+N$ and the screening pairing $\mu$.

Consider a two-dimensional $\mathcal N=(2,2)$ Landau-Ginzburg model with chiral fields $\phi^a$, $a=1,\ldots,n$, valued in a K\"ahler target space $Y$, with metric $g_{a\bar b}$ and holomorphic superpotential
\begin{equation}
    W:Y\longrightarrow\mathbb C.
\end{equation}
We assume that the critical points $v_i$ of $W$ are isolated and nondegenerate, and denote their critical values by $W_i=W(v_i)$. We normalize the energy of a static configuration as
\begin{equation}
    E=\int dx\left(g_{a\bar b}\,\partial_x\phi^a\,\partial_x\bar\phi^{\bar b} +\frac14 g^{a\bar b}\partial_a W\,\partial_{\bar b}\overline W\right).
\end{equation}
Completing the square gives
\begin{equation}
    E= \int dx\,\left|\partial_x\phi^a-\frac{e^{i\alpha}}{2} g^{a\bar b}\partial_{\bar b}\overline W\right|^2 +\operatorname{Re}\!\left(e^{-i\alpha}\Delta W\right),
\end{equation}
where $\Delta W=W(\phi(+\infty))-W(\phi(-\infty))$. The BPS bound $E\geq|\Delta W|$ is saturated for $\alpha=\arg\Delta W$ when
\begin{equation}
    \frac{d\phi^a}{dx}=\frac{e^{i\alpha}}{2} g^{a\bar b}\partial_{\bar b}\overline W.
\end{equation}
Solutions to this equation are the BPS solitons of the theory. A BPS soliton interpolating between vacua $v_i$ and $v_j$ has tension $T_{ij}=|W_j-W_i|$.\footnote{An alternative normalization that appears often in the literature drops the factor of $1/4$. In this case, the BPS equation has no factor of $1/2$ and the tension is $2|W_j-W_i|$.}

Along a BPS solution we have
\begin{equation}
    \frac{d}{dx}\left(e^{-i\alpha}W\right) =\frac12 g^{a\bar b}\partial_aW\,\partial_{\bar b}\overline W\geq0.
\end{equation}
Consequently, $\operatorname{Im}(e^{-i\alpha}W)$ is constant and $\operatorname{Re}(e^{-i\alpha}W)$ increases monotonically. The image of the soliton under $W$ is therefore the straight segment from $W_i$ to $W_j$.

This description has a useful geometric interpretation. Near a nondegenerate critical point, the superpotential takes the local form $W=W_\alpha+\frac12\sum_a(\phi^a)^2$. The corresponding vanishing cycle $\Delta_\alpha$ is an $(n-1)$-sphere, obtained by following the BPS gradient flow toward the critical point. The vanishing cycles can be transported to a common regular fiber of $W$, where their intersections give the signed soliton counts. Choosing a distinguished ordering of the vacua, we write
\begin{equation}
    N_{\alpha \beta}=\Delta_\alpha\circ\Delta_\beta\,,\qquad \alpha < \beta,
\end{equation}
with $N_{\alpha \beta}=0$ for $\alpha \geq \beta$. The Stokes matrix is
\begin{equation}
    S=\mathbf1+N.
\end{equation}
Because $\Delta_\alpha$ is a middle-dimensional cycle in a fiber of complex dimension $n-1$, its intersection form obeys
\begin{equation}
    \Delta_\alpha \circ\Delta_\beta = (-1)^{n-1}\Delta_\beta\circ\Delta_\alpha.
\end{equation}
With the sign convention used in section \ref{sec:SCREEN}, the full intersection pairing is
\begin{equation}
    \mu=S+(-1)^{n-1}S^T.
\end{equation}
Thus for $n$ odd the pairing is symmetric, while for $n$ even it is anti-symmetric. In particular,
\begin{equation}
    \mu=\begin{cases}
        S+S^T,&n\text{ odd},\\
        S-S^T,&n\text{ even}.
    \end{cases}
\end{equation}

Finally, the vanishing cycles may be extended along rays in the $W$-plane to form noncompact Lefschetz thimbles. These provide the A-type D-brane boundary conditions familiar from the LG description \cite{Hori:2000ck}. For our purposes, the relevant point is that the signed soliton data determine the pairing $\mu$, whose image gives the screening relations on the integral lattice $\Lambda$ of boundary states. In the notation of the main text,
\begin{equation}
    \mathbb B=\Lambda/\operatorname{im}(\mu)=\operatorname{coker}(\mu).
\end{equation}

\section{Solitons and Interfaces in LG Models} \label{app:LGinterface}

In this Appendix we discuss some further aspects of solitons in LG models with a supersymmetric interface.\footnote{For a more sophisticated discussion of related notions, see reference \cite{Gaiotto:2015aoa, Gaiotto:2015zna}.} Here we allow the interface to have finite thickness and localized degrees of freedom. We assume that we have two LG models $\mathcal{T}_1$ and $\mathcal{T}_{2}$ with corresponding superpotentials $W_{1}(\{ X_1 \})$ and $W_{2}(\{ X_2 \})$. To study soliton transmission through the interface, we can either work in this theory or the one obtained via folding, where we deal with the LG theory with superpotential $W \equiv W_{1} - W_{2}$.

As discussed in section \ref{sec:INTERFACE}, we can view the interface as a specific boundary state in the folded theory. In particular, given a basis of vacua $\vert \alpha, \overline{\beta} \rangle$ for the theory defined by $W = W_1 - W_2$, the interface amounts to a choice of state:
\begin{equation}
\vert \mathbf{I} \rangle \rangle = \underset{\alpha, \overline{\beta}}{\sum} \mathbf{I}_{\alpha \overline{\beta}} \vert \alpha, \overline{\beta} \rangle \in \Lambda_{1} \otimes \overline{\Lambda}_{2}.
\end{equation}
Strictly speaking, we do not have a soliton which interpolates from a fixed boundary state $\vert \alpha , \overline{\beta} \rangle$ to $\vert \mathbf{I} \rangle \rangle$, since generically the interface state is not a vacuum of the folded theory. Nevertheless, we can still study the spectrum of solitons
connecting vacua of the folded theory. This corresponds to the special case where $\vert \mathbf{I} \rangle \rangle = \vert \alpha^{\prime}, \overline{\beta^{\prime}} \rangle$, namely we specialize to a vacuum state. Now, in this special case, we can speak of a soliton which interpolates between the two vacua. This
defines a constant phase trajectory in the $W$-plane which passes from $W(\alpha, \beta) = W_1(\alpha) - W_{2}(\beta)$ to $W(\alpha^\prime, \beta^{\prime}) = W_{1}(\alpha^\prime) - W_{2}(\beta^{\prime})$. The central charge of the BPS soliton is proportional to the difference, so in a suitable normalization convention for the kinetic terms, we have:
\begin{equation}
Z_{\alpha \beta ; \alpha^{\prime} \beta^{\prime}} = W_{\alpha^\prime, \beta^\prime} - W_{\alpha, \beta} = (W_{\alpha^{\prime}} - W_{\alpha^{}}) + (W_{\beta} - W_{\beta^{\prime}}) = Z^{(1)}_{\alpha \alpha^\prime} - Z^{(2)}_{\beta \beta^{\prime}}.
\end{equation}

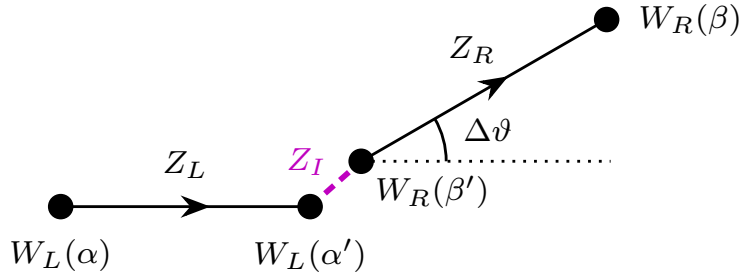
\begin{figure}
\centering
\scalebox{1.5}{
\begin{tikzpicture}[
  font=\small,
  mid arrow/.style={postaction={decorate}, decoration={markings,
    mark=at position 0.6 with {\arrow{Stealth[scale=1.2]}}}},
]

  \coordinate (A)  at (0,0);
  \coordinate (Ap) at (2.2,0);
  \coordinate (Bp) at (2.65,0.4);
  \coordinate (B)  at ($(Bp)+(30:2.5)$);

  \draw[DottedLine] (Bp) -- ++(2.2,0);
  \draw[ThickLine] ([shift={(0:0.75)}]Bp) arc[start angle=0, end angle=30, radius=0.75];
  \node at ([shift={(14:1.15)}]Bp) {\scriptsize{$\Delta\vartheta$}};

  \draw[ThickLine, mid arrow] (A) -- (Ap);
  \draw[DashedPurpleLine] (Ap) -- (Bp);
  \draw[ThickLine, mid arrow] (Bp) -- (B);

  \node[Circle] at (A)  {};
  \node[Circle] at (Ap) {};
  \node[Circle] at (Bp) {};
  \node[Circle] at (B)  {};

  \node[below=4pt] at (A)  {\scriptsize{$W_L(\alpha)$}};
  \node[below=4pt] at (Ap) {\scriptsize{$W_L(\alpha')$}};
  \node[anchor=north west, inner sep=2pt] at ([shift={(0.05,-0.05)}]Bp) {\scriptsize{$W_R(\beta')$}};
  \node[right=4pt] at (B)  {\scriptsize{$W_R(\beta)$}};
  \node[above=3pt] at ($(A)!0.5!(Ap)$) {\scriptsize{$Z_L$}};
  \node[anchor=south east, inner sep=2pt] at ($(Ap)!0.5!(Bp)$) {\color[RGB]{191,0,191}\scriptsize{$Z_I$}};
  \node[anchor=south east, inner sep=3pt] at ($(Bp)!0.6!(B)$) {\scriptsize{$Z_R$}};
\end{tikzpicture}}
\caption{Schematic depiction of the interface changing the trajectory of a soliton profile by enacting a deflection angle.}
\label{fig:DeflectionInterface}
\end{figure}

Next, consider the interpretation in the unfolded theory. The soliton profile in this case amounts to joining $\alpha \overset{S_1}{\rightarrow} \alpha^{\prime} \overset{\mathbf{I}}{\rightarrow} \beta^{\prime} \overset{S_2}{\rightarrow} \beta$. We can think of this as different BPS configurations which preserve different subsets of two real supercharges. Letting $Z_s = \vert Z_s \vert \exp(i \vartheta_s)$ with $s = 1,2,\mathbf{I}$ for the left soliton, right soliton and interface, we conclude that the interface carries central charge $Z_{I} = W_{2}(\beta^{\prime}) - W_{1}(\alpha^{\prime})$. This leads to a ``deflection angle'' which changes the trajectory. See figure \ref{fig:DeflectionInterface} for a depiction.

For interfaces constructed as a linear combination of such boundary states, the situation is clearly more challenging. Returning to:
\begin{equation}
\vert \mathbf{I} \rangle \rangle = \underset{\alpha, \overline{\beta}}{\sum} \mathbf{I}_{\alpha \overline{\beta}} \vert \alpha, \overline{\beta} \rangle \in \Lambda_{L} \otimes \overline{\Lambda}_{R},
\end{equation}
each summand specifies a classical soliton trajectory connecting different massive vacua. The linear combination of these different contributions complicates the interpretation since now there is no single trajectory to keep track of. Of course, when the different trajectories are clustered closely together, the same sort of semi-classical picture in terms of a single deflection angle still holds. It would be interesting to study this further.

\section{Compact Target Space Example} \label{app:FURTHER}

In this Appendix we verify that the BSSG is trivial for models with a compact target space. This is to be expected as compact models should not have any global symmetries. We check this explicitly for the $1^9$ Gepner model. Other Gepner models (and more generally any compact model), such as the K3 sigma model and the $2^6$ model, follow analogously.

The $1^9$ model consists of nine $\mathcal{A}$-type models subject to a diagonal $\mathbb{Z}_3$ orbifold. The superpotential is
\begin{equation}
    W=\frac{1}{3}\sum_{i=1}^9 Y_i^3, \qquad Y_i\mapsto \omega Y_i, \qquad \omega^3=1.
\end{equation}
The intersection pairing / index for this model has already been computed in \cite{Becker:2006ks, Becker:2024ijy}. We briefly review their computation and refer the reader to \cite{Becker:2006ks, Becker:2024ijy} for more details.

For a single constituent minimal model, the massive deformation gives two vacua connected by a soliton. The corresponding Stokes matrix is therefore
\begin{equation}
    S_1=\begin{pmatrix}
        1&1\\
        0&1
    \end{pmatrix}.
\end{equation}
The $\mathbb{Z}_3$ action on the charge lattice is given by
\begin{equation}
    H=-S_1^{-1}S_1^T.
\end{equation}
The Stokes matrix for the full $1^9$ model is obtained by taking the tensor product of that for the nine constituent minimal models: $S=S_1^{\otimes 9}$. The diagonal $\mathbb{Z}_3$ orbifold acts on the resulting charge lattice through $M=H^{\otimes 9}$. This gives the intersection pairing
\begin{equation}
    \mu=S_1^{\otimes 9} \left( \mathbf{1}+H^{\otimes 9} +(H^{\otimes 9})^2 \right).
\end{equation}
This naively gives a pairing matrix of size $512 \times 512$. However, as shown in \cite{Becker:2006ks}, the orbifold projection reduces the independent charge lattice from rank 512 to rank 170. In particular, the generators of the charge lattice can be labeled by binary vectors $\boldsymbol{a}=(a_1,\ldots,a_9)$ with $a_i\in\{0,1\}$. The first 170 such vectors, ordered as binary integers, form an integral basis for the charge lattice. The corresponding $170\times170$ intersection matrix $\mu_{\mathrm{orb}}$ has entries
\begin{equation}
    (\mu_{\mathrm{orb}})_{\boldsymbol{a}\boldsymbol{b}} =\sum_{k=0}^{2} \prod_{i=1}^{9}(S_1H^k)_{a_i b_i} =\mathbf{1}_{\boldsymbol{a}\leq\boldsymbol{b}} -\mathbf{1}_{\boldsymbol{b}\leq\boldsymbol{a}},
\end{equation}
where the inequalities are understood componentwise, and $\mathbf{1}_{P}$ is one when the condition $P$ is satisfied and zero otherwise.

This matrix is unimodular, so its Smith normal form is
\begin{equation}
    \operatorname{SNF}(\mu_{\mathrm{orb}}) =\operatorname{diag}( \underbrace{1,\ldots,1}_{170}).
\end{equation}
Consequently, the BSSG is
\begin{equation}
    \mathbb{B}_{1^9} =\operatorname{coker}(\mu_{\mathrm{orb}})=0.
\end{equation}
This agrees with the expectation that the BSSG of a compact model is trivial.

\newpage

\bibliographystyle{utphys}
\bibliography{SolitonScreeningGroup}

\end{document}